\documentclass[12pt]{iopart}

\usepackage{iopams}

\usepackage{graphicx} 
\graphicspath{{figures/}} 

\usepackage{booktabs} 
\usepackage[font=small,labelfont=bf]{caption} 
\usepackage{amsfonts, amsthm, amssymb} 
\usepackage{subcaption}
\usepackage{dcolumn}
\usepackage{xcolor}
\usepackage{hyperref}
\usepackage[normalem]{ulem}
\usepackage{systeme}

\usepackage{xcolor}
\usepackage{amssymb}
\begin{document}

\title[A new GRMHD code with neutrino leakage]{Spritz: General Relativistic Magnetohydrodynamics with Neutrinos}

\author{F. Cipolletta$^{1, \, 2, \, 3}$, J. V. Kalinani$^{4, \, 5}$, E. Giangrandi$^{6}$, B. Giacomazzo$^{6, \, 7, \, 8, \, \ast}$, R. Ciolfi$^{9, \, 5}$, L. Sala$^{6, \, 7}$\footnote{current affiliation: Universit\`a degli Studi di Trento, Dipartimento di Fisica, Via Sommarive 14, I-38123 Trento, Italy}, B. Giudici$^{6}$}

\address{
	$^{1}$Center for Computational Relativity and Gravitation, School of Mathematical Sciences, Rochester Institute of Technology, 85 Lomb Memorial Drive, Rochester, New York 14623, USA
	\\$^{2}$INFN-TIFPA, Trento Institute for Fundamental Physics and Applications, Via Sommarive 14, I-38123 Trento, Italy
	\\$^{3}$Universit\`a degli Studi di Trento, Dipartimento di Fisica, Via Sommarive 14, I-38123 Trento, Italy
	\\$^{4}$Universit\`a di Padova, Dipartimento di Fisica e Astronomia, Via Francesco Marzolo 8, I-35131 Padova, Italy
	\\$^{5}$INFN, Sezione di Padova, Via Francesco Marzolo 8, I-35131 Padova, Italy
	\\$^{6}$Universit\`a degli Studi di Milano - Bicocca, Dipartimento di Fisica G. Occhialini, Piazza della Scienza 3, I-20126 Milano, Italy
	\\$^{7}$INFN, Sezione di Milano-Bicocca, Piazza della Scienza 3, I-20126 Milano, Italy
	\\$^{8}$INAF, Osservatorio Astronomico di Brera, via E. Bianchi 46, I-23807 Merate (LC), Italy
	\\$^{9}$INAF, Osservatorio Astronomico di Padova, Vicolo dell'Osservatorio 5, I-35122 Padova, Italy
}

\ead{$^{\ast}$\texttt{bruno.giacomazzo@unimib.it}}
\vspace{10pt}
\begin{indented}
	\item[]December 2020
\end{indented}

\begin{abstract}
We here present a new version of the publicly available general relativistic magnetohydrodynamic (GRMHD) code \texttt{Spritz}, which now includes an approximate neutrino leakage scheme able to handle neutrino cooling and heating. 
The leakage scheme is based on the publicly available \texttt{ZelmaniLeak} code, with a few modifications in order to properly work with \texttt{Spritz}.
We discuss the involved equations, physical assumptions, and implemented numerical methods, along with a large battery of general relativistic tests performed with and without magnetic fields. Our tests demonstrate the correct implementation of the neutrino leakage scheme, paving the way for further improvements of our neutrino treatment and the first application to magnetized binary neutron star mergers. We also discuss the implementation in the \texttt{Spritz} code of high-order methods for a more accurate evolution of hydrodynamical quantities.
\end{abstract}

%
\vspace{2pc}
%
\submitto{\CQG}
%
%
%

\section{Introduction}

Binary neutron star (BNS) mergers are among the most powerful sources of gravitational waves (GWs) that can be detected by current ground-based GW detectors. The detection of GW170817~\cite{TheLIGOScientific:2017qsa} also confirmed that these systems may emit bright electromagnetic (EM) signals and, in particular, short gamma-ray bursts (GRBs) and 
kilonovae (e.g., \cite{Monitor:2017mdv,GBM:2017lvd,Troja2017,Margutti2017,Hallinan2017,Alexander2017,Mooley2018a,Lazzati2018,Lyman2018,Mooley2018b,Ghirlanda2019,Kasen2017,Pian2017,smartt2017kilonova}). In order to properly model the merger and post-merger evolution of these systems and thus establish a reliable connection with their multimessenger observations, one needs to account not only for general relativistic effects, but also for other key physical ingredients such as magnetic fields, a temperature and composition dependent equation of state describing the behaviour of matter, and neutrino emission and re-absorption. 
For instance, neutrino effects and magnetic fields are both crucial (i) to accurately model the BNS merger ejecta and their composition, which are in turn responsible for the kilonova emission and the associated heavy element nucleosynthesis (e.g., \cite{Foucart:2016rxm,Radice:2018pdn,Ciolfi2020b,CiolfiKalinani2020} and refs.~therein), and (ii) in the context of short GRB jet formation, where magnetic fields are most likely the main driver (e.g., \cite{Just2016,Perego2017,Ruiz2016,Ciolfi2020a}) while neutrino radiation may play an important role in altering the baryon pollution along the spin axis of the remnant, which in turn may affect the successful propagation of the corresponding outflow (e.g., \cite{Moesta2020}). 
Including all of the above effects in one code is however rather challenging and only very few magnetized BNS merger simulations with neutrino treatment (via an approximate leakage scheme) have been presented so far \cite{palenzuela2015effects, Most2019}. 

Here, we present a new publicly available version of our general relativistic magnetohydrodynamic (GRMHD) code named \texttt{Spritz} \cite{cipolletta2020spritz,spritz}, based on the \texttt{EinsteinToolkit} infrastructure~\cite{loffler2012einstein, zilhao2013introduction, etk}. 
This new version of \texttt{Spritz} can handle finite temperature tabulated equations of state (EOSs) as well as neutrino cooling/heating along with magnetic fields. 
In particular, the neutrino treatment is built around the \texttt{ZelmaniLeak} code~\cite{stellarcollapse}, implementing a ray-by-ray neutrino leakage scheme. \texttt{ZelmaniLeak} has already been employed in the context of BNS mergers and in particular in GRMHD simulations starting from a non-magnetized post-merger system to which a magnetic field is added by hand \cite{Moesta2020}. We note that while more advanced schemes have been discussed in the literature (e.g., \cite{Foucart2020}), only simple leakage schemes have been so far employed to study merging BNSs with both magnetic fields and neutrinos \cite{palenzuela2015effects, Most2019} (and the corresponding implementations are not publicly available).
Therefore, neutrino leakage represents a natural starting point for the inclusion of this key physical ingredient in \texttt{Spritz}.

During the writing of this paper we also finished implementing in the code new high-order methods that are described in~\ref{highord}, where we show that the code can now reach, in some scenarios, fifth-order convergence. High-order methods have been shown in the literature to be very important in order to obtain accurate GW signals and a better description of the matter dynamics (e.g., see~\cite{Radice2014, bernuzzi2016waveforms}). At the time of writing, only few other GRMHD codes for BNS simulations employ high-order methods~\cite{Most2019,Moesta2020,Aguilera2020}.
The new version of the \texttt{Spritz} code can be found on Zenodo as version 1.1.0~\cite{spritz}.

The paper is organized as follows. In Section~\ref{equations}, we present the equations and assumptions behind the adopted neutrino leakage scheme. 
Section~\ref{methods} provides an overview of the new numerical methods included in the \texttt{Spritz} code, from tabulated EOS handling and conservative-to-primitive recovery to the neutrino leakage implementation. Section~\ref{tests} is devoted to a large set of tests, through which we validate the novel features of  \texttt{Spritz}. Finally, we summarize our results in Section~\ref{conclusions}.

We use geometric units such that $G=c=M_\odot=1$ unless specified otherwise. Greek indices go from 0 to 3, Latin indices from 1 to 3, and summation over repeated indices is assumed. As usual, we employ a $(-\, +\, +\, +)$ metric signature. We use a $3+1$ decomposition of the space-time, where the 4-metric is indicated with $g_{\mu \nu}$ and $ds^{2} = g_{\mu \nu} dx^{\mu} dx^{\nu} = -\left( \alpha^{2} - \beta^{i}\beta_{i} \right) dt^2 + 2 \beta_{i} dx^{i} dt + \gamma_{ij} dx^{i}dx^{j}$. $\alpha$ is the lapse function, $\beta^i$ is the shift vector, and $\gamma_{ij}$ is the 3-metric. Moreover $g$ and $\gamma$ represent the determinant of $g_{\mu \nu}$ and $\gamma_{ij}$ respectively.

\section{Basic Equations and Assumptions\label{equations}}

In the present Section we discuss the equations that are solved by the new version of our GRMHD code which now include also the contribution of neutrino emission and absorption. We will mainly focus on the new additions to the code and refer the reader to our previous paper for more details on the equations and methods used to solve the GRMHD equations~\cite{cipolletta2020spritz}. We remind the reader that the (Eulerian) magnetic field $B^i$ is evolved via a staggered-vector-potential formulation. The equations for the evolution of the rest-mass density $\rho$, three-velocity $v^i$, and specific internal energy $\varepsilon$ are set according to the following conservative formulation:
\begin{equation}
	\label{consequat}
	\frac{1}{\sqrt{-g}} \left[ \partial_t \left( \sqrt{\gamma} \bi{F}^0 \right) + \partial_i \left( \sqrt{-g} \bi{F}^i \right) \right] = \bi{S}^i \, ,
\end{equation}
being $\bi{F}^0 \equiv \left[ D, S_j, \tilde{\tau} \right]$\footnote{ We use the symbol $\tilde{\tau}$ instead of the commonly used $\tau$ to avoid confusion with the optical depth $\tau$ used later in the paper.} the vector of conserved variables, defined in terms of the primitive ones as
\begin{equation}
	\label{P2Csystem}
	\eqalign{
		D &\equiv \rho W, \\
		S_{j} &\equiv \left( \rho h + b^2 \right) W^2 v_{j} - \alpha b^0 b_{j}, \\
		\tilde{\tau} &\equiv \left( \rho h + b^2 \right) W^2 - \left( P + P_\mathrm{mag} \right) - \alpha^2 \left( b^0 \right)^2 - D \, ,
	}
\end{equation}
where $W=1/\sqrt{1-v^2}$ is the Lorentz factor, $P$  is the gas pressure, $h=1+\varepsilon+P/\rho$ is the relativistic specific enthalpy, $P_\mathrm{mag}=b^2/2$ is the magnetic pressure, $b^0 = (W B^i v_i)/{\alpha}$, $b^i = (B^i + \alpha b^0 u^i)/{W}$, $b^2 \equiv b^\mu b_\mu = \left[B^2 + \alpha^2 \left( b^0 \right)^2\right]/{W^2}$, $B^2=B^i B_i$, and $u^\mu$ is the fluid four-velocity. $\bi{F}^i$ is instead the vector of fluxes defined as
\begin{equation}
	\label{conservedvector}
	\bi{F}^i \equiv \left[ \eqalign{
		\qquad\qquad D\tilde{v}^i / \alpha \\
		\; S_j \tilde{v}^i / \alpha + \left( P + P_\mathrm{mag} \right) \delta^i_j - b_j B^i / W \\
		\; \tilde{\tau} \tilde{v}^i / \alpha + \left( P + P_\mathrm{mag} \right) v^i - \alpha b^0 B^i / W 
	} \right] \, ,
\end{equation}
where $\tilde{v}^i \equiv \alpha v^i - \beta ^i$ and $\beta^i$ is the shift, while $\bi{S}^i$ the vector of sources that reads
\begin{equation}
	\label{sources}
	\bi{S}^i \equiv \left[ \eqalign{
		\qquad\qquad 0 \\
		\;\; T^{\mu \nu} \left( \partial_{\mu} g_{\nu j} - \Gamma^{\delta}_{\nu \mu} g_{\delta j} \right) \\
		\; \alpha \left( T^{\mu 0} \partial_{\mu} \ln{\alpha} - T^{\mu \nu} \Gamma^{0}_{\nu \mu} \right) 
	} \right]\, ,
\end{equation}
where $T^{\mu \nu}$ is the energy-momentum tensor, given by $T^{\mu \nu} = \left( \rho h + b^2 \right) u^{\mu} u^{\nu} + \left( P + P_\mathrm{mag} \right) g^{\mu \nu} - b^{\mu} b^{\nu}$, and $\Gamma^{\sigma}_{\nu \mu}$ are the Christoffel symbols defined from the 4-metric $g_{\mu \nu}$.

We note that the above equations do not include the contribution of neutrino emission and reabsorption. Following an operator-split approach, the GRMHD evolution step is first performed without such contribution and then the neutrino problem is solved via the leakage scheme. Finally, the variables $Y_e$ and $\varepsilon$ are updated accordingly, thus including the effects of neutrino radiation on the GRMHD evolution itself (see Sections \ref{nueqs} and \ref{nuimpl}).

\subsection{Electron Fraction}
\label{elefrac}

In order to properly include neutrino emission and absorption, we need to add one evolution equation for the electron fraction, which we define as
\begin{equation}
\label{yedef}
Y_e = \frac{n_e}{n_p + n_n} \, ,
\end{equation}
being $n_e$, $n_p$, and $n_n$ the electron, proton, and neutron number densities.

From the local conservation of the total baryon number, neglecting the mass difference between neutrons and protons, we obtain the following equation for the electron fraction, valid in absence of neutrino emission/absorption:
\begin{equation}
\label{leptcons}
\nabla_{\mu} \left( Y_e \rho u^{\mu} \right) = 0 \, ,
\end{equation}
expressing the fact that $Y_e$ is advected along the fluid lines. This equation is commonly referred to as the \textit{electron fraction advection} and can be expressed in a hyperbolic conservative form as
\begin{equation}
\label{hypconsyeeq}
\partial_t \left( \sqrt{\gamma} D Y_e \right) + \partial_i \left[ \alpha \sqrt{\gamma} D Y_e \left( v^i - \frac{\beta^i}{\alpha} \right) \right] = 0 \, .
\end{equation}

In presence of reactions involving neutrinos, the local electron fraction obtained from the above equation is then modified according to Equation~\ref{Ye_mod} (see Section \ref{nueqs}).

\subsection{Equation of State}
\label{eos}

The \texttt{Spritz} code can handle tabulated finite-temperature and composition dependent EOS via the \texttt{EOS\_Omni} thorn included in the Einstein Toolkit. This is crucial since a proper description of the matter composition depending on temperature is necessary in order to estimate the emission and absorption rates associated with the different processes involving neutrinos (see the next Section).
Moreover, as a consequence of such processes, $Y_e$ necessarily undergoes changes that must be estimated accurately when dealing with dynamical scenarios. 

The exact matter composition at the typical densities reached in the core of an NS is still unknown and so is the correct EOS. A large number of proposed tabulated EOS inspired by nuclear physics calculations can be found in the literature (see, e.g., the database in \cite{stellarcollapse} and \cite{compose} for several examples). These EOS are usually three-dimensional tables where every hydrodynamical variable, such as the gas pressure $P$ or the specific internal energy $\varepsilon$, can be related to the rest-mass density $\rho$, the temperature $T$, and the electron fraction $Y_e$.

When building initial data, however, a one-dimensional (i.e., barotropic) EOS is typically needed, where $P$ is just a function of $\rho$. In this case, reducing the three-dimensional table $P=P(\rho, T, Y_e)$ to a simpler one-dimensional relation $P=P(\rho)$ becomes necessary, implying that two conditions on the NS matter should be imposed. 
The first and most common one is to assume the NS to be initially in $\beta$-equilibrium, which is a reasonable assumption for old NSs, such as those encountered in BNS or NSBH binary systems prior to merger. As a second assumption, one may decide to fix either a constant value for the entropy (\textit{S--slicing} condition) or for the temperature (\textit{T--slicing} condition). The latter is the one typically used in BNS or NSBH merger simulations since it is reasonable to expect NSs to be cold prior to merger. In this paper, along with the standard \textit{T--slicing} condition, we have also used the \textit{S--slicing} condition to test the ability of our code in dealing with ``hot'' NSs.

All the computations presented in this paper are performed adopting the LS220 EOS \cite{lattimer1991generalized}, that has been already used in a number of papers dealing with the evolution of BNS systems (e.g., \cite{Foucart:2016rxm,Radice++2018,Bernuzzi2020}).

\subsection{Neutrino Emission and Absorption}
\label{nueqs}

During the merger of BNS or NSBH systems, temperatures as high as $T \sim 10 \,\mathrm{MeV}\sim 10^{11}$\,K can be produced and also the electron fraction $Y_e$ may change considerably. In this scenario, neutrinos play a key role in both the transport of energy and in determining the evolution of $Y_e$ and temperature, which are in turn crucial parameters for the $r$-process nucleosynthesis taking place in the ejected matter and the subsequent production of heavy elements. 
A proper estimate of the rates of the different reactions involving neutrinos is thus necessary in order to compute the nucleosythesis yields and to model the radiactively-powered kilonova signals accompanying such mergers (as the one already observed after GW170817; e.g.,~\cite{Pian2017,smartt2017kilonova}).

The typical timescale for weak processes producing neutrinos can be estimated from the changing electron fraction as
\begin{equation}
\label{tsWP}
t_{\rm WP} \sim \left| \frac{Y_e}{\dot{Y}_e} \right| \ll t_{\rm dyn}\, ,
\end{equation}
being $t_{\rm dyn}$ the dynamical timescale of the simulated astrophysical event \cite{sekiguchi2010stellar}. 
By carrying away energy, neutrinos can significantly cool down the (meta)stable NS remnant of a BNS merger or the accretion disk around the spinning BH resulting from either a BNS or an NSBH merger (e.g., \cite{Foucart:2016rxm}). 
Moreover, a fraction of the emitted neutrinos may be reabsorbed by the outer material, inducing heating and leptonization of the material itself.
The surface where the neutrino optical depth is $\tau = 2/3$ conventionally defines the ``neutrinosphere'' (e.g., \cite{rosso2018introduction}), which separates the diffusive regime of the high-density interiors ($\,\gtrsim10^{12}$\,g cm$^{-3}$; e.g., \cite{ruffert1997coalescing}) and the nearly free streaming regime of the exterior. The intermediate region between $\tau \ll 1$ and $\tau\gg 1$ (i.e.~where neutrinos are neither free to escape nor fully trapped) is the challenging one for neutrino transport. 
In its energy averaged version, the optical depth along each path $\xi$ followed by neutrinos can be defined as \cite{o2010new}
\begin{equation}
\label{optdep}
\tau_\xi = \int_\xi \rho(\mathbf{x}) k(\mathbf{x}) \sqrt{\gamma_{ij} dx^i dx^j}\, ,
\end{equation}
being $k(\mathbf{x})$ the energy averaged opacity at position $\mathbf{x}$. 
The path giving the minimum optical depth is the favoured one for neutrino escape and allows us to define a single optical depth for each given location
\begin{equation}
\label{pointoptdep}
\tau(\mathbf{x}) = \min\limits_{\xi \in \Xi} \tau_\xi = \min\limits_{\xi \in \Xi} \int_\xi \rho(\mathbf{x}) k(\mathbf{x}) \sqrt{\gamma_{ij} dx^i dx^j},
\end{equation}
where $\Xi$ is the set of all possible paths including position $\mathbf{x}$.

The complexity and extremely high computational cost of the full neutrino transport problem solved via the Boltzmann radiation transport equations forced the introduction of approximate schemes and simplifying assumptions (e.g., \cite{Foucart2020} and refs.~therein).
We consider here a so-called neutrino leakage scheme, already employed successfully in BNS and NSBH simulations (e.g., \cite{sekiguchi2010stellar, deaton2013black, foucart2014neutron, palenzuela2015effects,radice2016dynamical}). 
In particular, we adopt the leakage method presented in \cite{o2010new,ott2013general}, which has been implemented in the publicly available \texttt{ZelmaniLeak} code~\cite{stellarcollapse}. 
In what follows, we introduce the leakage scheme and the basic physical assumptions. 
The numerical implementation is instead discussed in the next Section (and in particular in \ref{nuimpl}).

In the neutrino leakage scheme adopted in this work, we consider three neutrino species, electron neutrino $\nu_e$, electron antineutrino $\bar{\nu}_e$, and heavy-lepton neutrinos $\nu_x$ (including $\nu_\mu$, $\bar{\nu}_\mu$, $\nu_\tau$, $\bar{\nu}_\tau$), and for each one we compute the local number and energy emission rates according to the following steps. 

The neutrino optical depths, which are crucial to determine the emission rates (see below), are computed under the assumption that neutrinos escape along radial paths from the center (ray-by-ray approach). For each species, we compute the local spectral averaged opacity as the sum of the opacities due to the scattering off nucleons, neutrino-nucleus scattering, and neutrino absorption by free nucleons (see \cite{rosswog2003high} for details). Then, we use these mean opacities to compute the optical depths along each radial path (Eq.~\ref{optdep}).

In the diffusive regime, the number and energy rates (i.e.~number and energy per unit volume, per unit time) can be written as \cite{rosswog2003high}
\begin{equation}
\label{Rdiff}
R^{\rm{diff}}_{\nu_i} = \frac{4 \pi c g_{\nu_i}}{(hc)^3} \frac{\zeta_{\nu_i}}{3 \chi^2_{\nu_i}} T F_0(\eta_{\nu_i}) \, ,
\end{equation}
\begin{equation}
\label{Qdiff}
Q^{\rm{diff}}_{\nu_i} = \frac{4 \pi c g_{\nu_i}}{(hc)^3} \frac{\zeta_{\nu_i}}{3 \chi^2_{\nu_i}} T^2 F_1(\eta_{\nu_i}) \, ,
\end{equation}
where $i=1,2,3$ and $\nu_1=\nu_e$,  $\nu_2=\bar{\nu}_e$, $\nu_3=\nu_x$, while $g_{\nu_1} = g_{\nu_2} = 1$ and $g_{\nu_3} = 4$. Moreover, $\zeta=(E^2\lambda)^{-1}$, $\chi=\tau/E^2$, with $E$ the average neutrino energy (computed assuming a Fermi-Dirac distribution at the local temperature $T$)  and $\lambda$ the mean free path, and $F_0(\eta)$, $F_1(\eta)$ are the Fermi integrals defined in \cite{takahashi1978beta} as function of the neutrino chemical potential $\eta$. 
Energy and number rates are also computed for the free neutrino emission regime ($Q^{\rm{free}}_{\nu_i}$ and $R^{\rm{free}}_{\nu_i}$), taking into account capture processes, electron-positron pair annihilation, plasmon decay, and nucleon-nucleon bremsstrahlung (see \cite{ott2013general, rosswog2003high}).
Finally, the actual emission rates are found by combining the free emission and diffusive ones as follows 
\begin{equation}
\label{Reff}
R^{\rm{eff}}_{\nu_i} = R^{\rm{free}}_{\nu_i} \left( 1 + \frac{R^{\rm{free}}_{\nu_i}}{R^{\rm{diff}}_{\nu_i}} \right),
\end{equation}
\begin{equation}
\label{Qeff}
Q^{\rm{eff}}_{\nu_i} = Q^{\rm{free}}_{\nu_i} \left( 1 + \frac{Q^{\rm{free}}_{\nu_i}}{Q^{\rm{diff}}_{\nu_i}} \right).
\end{equation}

For a given radial direction ($\theta,\phi$), the isotropic-equivalent neutrino luminosity incoming from below at a distance $r$ can be computed (in the coordinate frame) as 
\begin{equation}
\label{LCF}
\eqalign{
L^{\rm{iso}}_{\nu_i}(r,\theta,\phi) = 4 \pi \int^r_0 & \left[ \frac{\alpha(r',\theta,\phi)}{\alpha(r,\theta,\phi)} \right] Q^{\rm{eff}}_{\nu_i}(r',\theta,\phi) \alpha(r',\theta,\phi) W(r',\theta,\phi) \cr 
& \times\left[ 1 + v^r(r',\theta,\phi) \right] \sqrt{g_{rr}(r',\theta,\phi)} {r'}^2 dr' \, ,
}
\end{equation}
being $v^r$ the radial velocity.\footnote{Note that this expression neglects the time-of-flight of neutrinos, i.e.~it just collects together neutrinos emitted at a given time and at different radial locations. However, this is only used in the region where neutrino reabsorption is relevant and in the post-merger phase of a BNS or NSBH coalescence the extension of such a region is characterized by a light travel time much shorter than the timescale for a significant change in neutrino luminosities.}
We can also define a fluid rest frame (FRF) luminosity as
\begin{equation}
\label{LFRF}
L^{\rm{iso,FRF}}_{\nu_i} (r) = \frac{L^{\rm{iso}}_{\nu_i} (r)}{\alpha(r) W(r) \left[ 1 + v^r(r) \right]} \,\, .
\end{equation}
The heating and leptonization due to the reabsorption of a fraction of neutrinos by the material along their path (i.e.~$\nu_e$ and $\bar{\nu}_e$ reabsorption on neutrons and protons, respectively) is taken into account via the local heating rate \cite{ott2013general}
\begin{equation}
\label{qheat}
Q^{\rm heat}_{(\nu_e,\bar{\nu}_e)} = f_{\rm heat} \frac{L^{\rm{iso,FRF}}_{(\nu_e,\bar{\nu}_e)}}{4 \pi r^2} \sigma^{{\rm heat}}_{(\nu_e,\bar{\nu}_e)}  \frac{\rho}{m_{(n,p)}} X_{(n,p)} \left( 4.275 \tau_{(\nu_e,\bar{\nu}_e)} + 1.15 \right) e^{-2 \tau_{(\nu_e,\bar{\nu}_e)}} \, ,
\end{equation}
where $f_{\rm heat}$ is a scaling factor of order one (we set $f_{\rm heat}=1$), $\sigma^{{\rm heat}}_{(\nu_e,\bar{\nu}_e)}$ is the reabsorption cross-section (see below), $m_{(n,p)}$ and $X_{(n,p)}$ are the neutron or proton masses and mass fractions, and the factor $e^{-2 \tau_{(\nu_e,\bar{\nu}_e)}}$ is added to suppress heating at very large optical depths.
For the reabsorption cross-section, we adopt the following expression \cite{o2010new}
\begin{equation}
\label{sigmaheat}
\sigma^{\rm heat}_{(\nu_e,\bar{\nu}_e)} = \frac{1+3\alpha_{\rm{EC}}^2}{4} \sigma_0 \frac{\langle E^2 \rangle^{NS}_{(\nu_e,\bar{\nu}_e)}}{{(m_e c^2)^2}} \langle 1 - f_{(e^{\small {-}},e^{\small {+}})} \rangle,
\end{equation}
where $\alpha_{EC}=-1.25$,
$\sigma_0=1.76\times10^{-44}$\,cm$^2$,
$\langle E^2 \rangle^{NS}$ is the mean squared neutrino energy at the neutrinosphere, and $\langle 1 - f_{(e^{\small {-}},e^{\small {+}})} \rangle$ are the blocking factors defined in \cite{ruffert1996coalescing}.

The full neutrino emission and reabsorption problem at a given time is solved along each radial direction by moving outwards from the center and, at each radius, subtracting the heating rate from the emission rate, i.e.~$Q^{\rm{eff}}_{\nu_i} \rightarrow Q^{\rm{eff}}_{\nu_i}-Q^{\rm heat}_{\nu_i}$ and $R^{\rm{eff}}_{\nu_i} \rightarrow R^{\rm{eff}}_{\nu_i}-Q^{\rm heat}_{\nu_i}/\langle E \rangle_{\nu_i}^{NS}$, with $\langle E \rangle_{\nu_i}^{NS}$ the average neutrino energy at the neutrinosphere and $Q^{\rm heat}_{\nu_x}=0$.\footnote{As pointed out in \cite{o2010new}, the present gray heating scheme does not provide a perfect balance between emission and absorption, which would require a self-consistent radiation transport treatment.}

In order to couple the result to the GRMHD evolution, the $Y_e$ and $\varepsilon$ are then modified as follows:
\begin{equation}
	\label{Ye_mod}
	Y_e \rightarrow Y_e + \Delta t \frac{\partial Y_e}{\partial t} \, ,
\end{equation}
being $\Delta t$ the local time step, and where
\begin{equation}
	\frac{\partial Y_e}{\partial t} = \frac{R_{\bar{\nu_e}}^{\rm{eff}}-R_{\nu_e}^{\rm{eff}}}{\rho} m_n \, ,
\end{equation}
being $m_n$ the rest-mass of the neutron, and
\begin{equation}
	\label{eps_mod}
	\varepsilon \rightarrow \varepsilon+ \Delta t \frac{\partial \varepsilon}{\partial t} \, ,
\end{equation}
where
\begin{equation}
	\frac{\partial \varepsilon}{\partial t} = -\frac{\Sigma_i Q_{\nu_i}^{\rm{eff}}}{\rho} \, .
\end{equation}
%

\section{Numerical Methods\label{methods}}

The \texttt{Spritz} code makes use of the \texttt{EinsteinToolkit} framework. Details of the numerical methods used to solve the GRMHD equations are provided in~\cite{cipolletta2020spritz} and here we focus on the new parts of the code that handle the use of tabulated EOS and neutrino emission and absorption.
All the simulations reported in this paper use the \texttt{MacLachlan} thorn to evolve the spacetime in the BSSNOK formalism and the \texttt{Carpet} driver for adaptive mesh refinement (AMR).

\subsection{Equation of State Driver}
\label{EOSdriv}

As already stated in \Sref{eos}, the \texttt{Spritz} code adopts the \texttt{EOS\_Omni} thorn of the \texttt{EinsteinToolkit} software infrastructure. This thorn is able to handle a large variety of EOS, including ideal fluid, polytropic, and tabulated ones.

During our first tests with the \texttt{EOS\_Omni} thorn and tabulated EOS, we noticed that the \texttt{EOS\_Omni} thorn presented some limitations in dealing with such EOS type. In particular, to compute the temperature $T$ from the specific internal energy $\varepsilon$, the thorn adopts a Newton-Raphson routine with a fall-back to a bisection routine in case of too many iterations, after verifying that the root is bracketed. We found this algorithm to be not robust enough in cases when $T$ weakly depends on $\varepsilon$, which may lead $T$ to go out of the bounds present in the chosen table (see \cite{galeazzi2013implementation}). This problem was present in particular when dealing with NS initial data using the lowest $T$ available in the EOS table. Such initial data undergo a sharp temperature increase in the core of the NS due to numerical readjustment of the initial data given by the solution of the TOV equations. We proposed a modification of the \texttt{EOS\_Omni} thorn to the \texttt{EinsteinToolkit} developers that consisted in preferring the fall-back to the more robust bisection method in such cases. In this way, we verified the temperature $T$ to be always contained in the range available in the table. This modification was accepted and it is now included in the publicly available \texttt{EinsteinToolkit} since May 2020~\cite{ETKTuring}.

We performed all the simulations discussed in \Sref{tests} using this new version of the \texttt{EOS\_Omni} thorn. We therefore caution the reader that the \texttt{Spritz} code should be used with the May 2020 release of the \texttt{EinsteinToolkit} (or later versions) when using tabulated EOS.

\subsection{Initial Data}
\label{idprod}

In order to compute the initial data, one needs to reduce the 3D EOS table to a 1D EOS, in which the pressure $P$ is only a function of the rest-mass density $\rho$. To do this, we assume $\beta$-equilibrium and then apply the S--slicing or T--slicing condition mentioned in \Sref{eos}. We coded a python script for this purpose (available with the public version of \texttt{Spritz}) that produces a 1D tabulated EOS starting from a 3D tabulated EOS in \texttt{.h5} format, such as the ones provided in \cite{stellarcollapse}. The 1D EOS is saved in the \texttt{CompOSE} format~\cite{compose} that can be easily used with \texttt{LORENE}~\cite{lorene}. The initial data used in this paper, reproducing a single non-rotating NS (TOV), were in particular produced with the code \texttt{Nrotstar} that can compute equilibrium solutions for non-rotating or uniformly rotating NSs. These solutions are non-magnetized, but a magnetic field can be easily added to the initial data as long as the field strength is $\lesssim 10^{17}$\,G, such that no significant effects on the NS structure nor significant violations of the constraint equations are introduced.

To read the initial data in the \texttt{Spritz} code we developed the \texttt{ID\_Nrotstar} thorn which is simply a reader that makes use of the \texttt{LORENE} library to read initial data produced with \texttt{Nrotstar} and import them in the Cartesian grid used by the code. Since the initial data were produced assuming $\beta$-equilibrium, we also developed an additional thorn, \texttt{Spritz\_SetBeta}, that instead makes sure that, when computing the conservative variables from the primitive ones at iteration 0, the code uses the same 1D EOS used to compute the initial data. After the initial data are correctly imported and conserved variables computed, the evolution starts and the full 3D EOS table is used.

Both \texttt{ID\_Nrotstar} and \texttt{Spritz\_SetBeta} are part of version 1.1.0 of the \texttt{Spritz} code~\cite{spritz}.

\subsection{Conservative-to-Primitive Inversion}
\label{C2P}

When using tabulated EOS we employ the 1D method for the conservative-to-primitive inversion presented by Palenzuela et al. \cite{palenzuela2015effects}. This method is a modification to the 1D method already used in GRHD \cite{galeazzi2013implementation}. It consists of rewriting the conserved variables in the following way
\begin{equation}
\label{consrewr}
q \equiv  \frac{\tilde{\tau}}{D},\,
r \equiv  \frac{S^2}{D^2},\,
s \equiv  \frac{B^2}{D},\,
t \equiv  \frac{B_i S^i}{D^{\frac{3}{2}}},
\end{equation}
and searching for the independent variable $x \equiv hW$. One then looks for the solution of $f(x)=0$, where $f(x)=x-hW$. We point the reader to \cite{palenzuela2015effects} for more details about the algorithm. Here, it is only important to note that the Brent's method \cite{brent2013algorithms} is used for the root finding, where the independent variable $x$ should be properly bracketed, thus $x \in \left] x_{\rm L}, x_{\rm R} \right[ $, with $f \left( x_{\rm L} \right) \cdot f \left( x_{\rm R} \right) < 0$. 
The left and right bounds can be defined in the following way (see \cite{siegel2018recovery}):
\begin{equation}
\label{rootbounds}
\eqalign{
x_{\rm L} =& 1 + q - s, \\
x_{\rm R} =& 2 + 2q - s.
}
\end{equation}
If no consistent bound is found, then the point is set to atmosphere.

As we will show in \Sref{tests}, we are also interested in performing simulations where the initial temperature $T$ is forced to be constant. This may be useful in order to avoid spurious neutrino production in particular scenarios, e.g.~during BNS inspiral (for some examples, see \cite{kastaun2016structure, kastaun2017structure, martin2018role}) or when evolving a single cold NS (that may undergo a sharp initial rise of temperature as already mentioned in \Sref{EOSdriv}). However, the aforementioned 1D conservative-to-primitive scheme cannot be used in such cases and we adopt a modification of the \texttt{3eqs} method that was already implemented in the \texttt{Spritz} code (see \cite{cipolletta2020spritz} and \cite{giacomazzo2007whiskymhd} for details), where the $\tilde{\tau}$ variable is not used in computing the primitive variables. In particular, we refer to Eq.~(45) of \cite{giacomazzo2007whiskymhd}, defining the function
\begin{equation}
\label{fconstr}
f \left( W_{\rm guess} \right) \equiv S^2 - \left[ \left( \hat{Z} + B^2 \right)^2 \frac{W_{\rm guess}^2 -1}{W_{\rm guess}^2} -\frac{2 \hat{Z} + B^2}{\hat{Z}^2} \left( B^i S_i \right)^2 \right]\, ,
\end{equation}
where
\begin{equation}
\label{zhat}
\hat{Z} = W_{\rm guess}^2 \left( \hat{\rho} + \hat{\rho} \hat{\varepsilon} + \hat{P} \right)\,,
\end{equation}
\begin{equation}
\label{rhohat}
\hat{\rho} = \frac{D}{W_{\rm guess}}\, ,
\end{equation}
and $\hat{P}$ and $\hat{\varepsilon}$ can be computed via the EOS using $\hat{\rho}$ and the constrained value of $T$. 
The algorithm proceeds as follows:
\begin{enumerate}
\item the initial guess for the solution is assumed to be $W_{\rm guess} \in [ 1.0,1.5 ]$\footnote{This corresponds to assuming $v \in [ 0.0, 0.75 c ]$};
\item $\hat{\rho}$, $\hat{P}$, $\hat{\varepsilon}$, and $\hat{Z}$ are computed using the EOS with the constrained value of $T$ and the conserved variables;
\item if, using \Eref{fconstr}, $f(1) \cdot f(1.5) > 0$, the point is actually set to the atmosphere;
\item the Brent's method \cite{brent2013algorithms} is applied to the function $f$ defined in \Eref{fconstr}.
\end{enumerate}

We note that we use Eq.~(45) but not Eq.~(46) of \cite{giacomazzo2007whiskymhd} when forcing the temperature to be constant. Therefore, we also need to update the value of $\tilde{\tau}$ after each conservative-to-primitive calculation in order to guarantee consistency between primitive and conservative variables. This is similar to what is done in other codes when using a cold EOS during the evolution. As we will show in \Sref{tests}, the code is able to easily switch from a constrained to a free temperature evolution without particular problems.

\subsection{Neutrino Leakage Implementation}
\label{nuimpl}

Our implementation of the neutrino leakage scheme described in Section \ref{nueqs} is based on the thorn \texttt{ZelmaniLeak} available at the \texttt{stellarcollapse} website \cite{stellarcollapse} and firstly presented in \cite{o2010new}. In particular, we employ version \texttt{20161117} of such thorn. The thorn \texttt{ZelmaniLeak} uses all the cross-sections and heating rate described in section~\ref{nueqs} and these cannot be modified by the user unless the code itself is modified. Nevetheless, the user can choose whether to activate neutrino heating or not as well as to include or not neutrino emission since the beginning of the simulation or after some time. Moreover, the user can freely set the number of radii across which the optical depth is computed.

\section{Tests}
\label{tests}

\begin{figure}[t!]
\begin{center}
\includegraphics[width=0.45\linewidth]{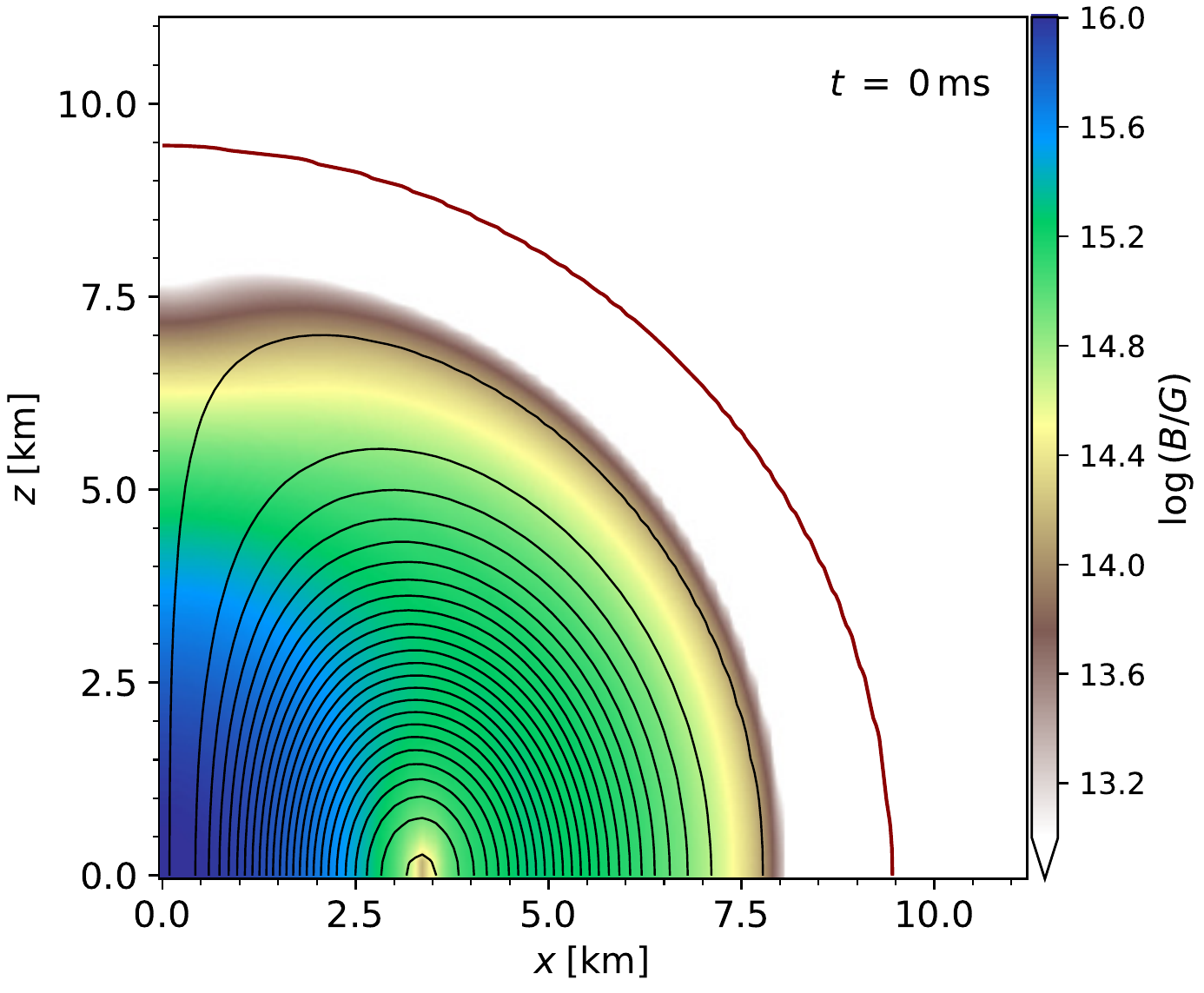}
\includegraphics[width=0.45\linewidth]{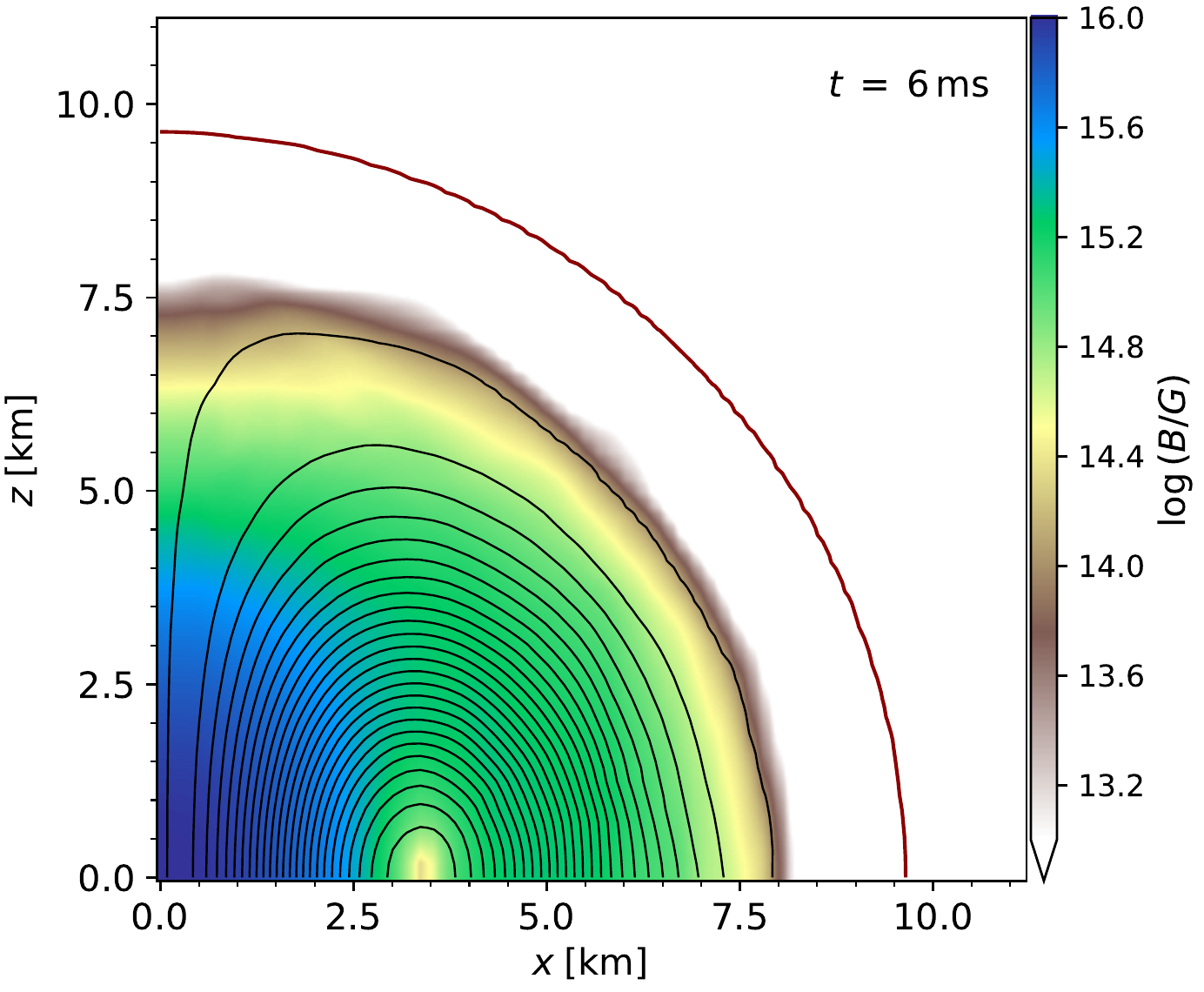}
\caption{\label{FigBfield2D} Left panel: initial magnetic field setup for simulations 13 and 14 (see Table \ref{table:IDB15}). Black lines represent isocontours of the $\phi$-component of the vector potential, while the red line corresponds to $\rho\simeq3\times10^{12}$\,g/cm$^{3}$. Right panel: as in the left panel but at the end of simulation 16.}
\end{center}
\end{figure}

In this Section, we report the full set of tests that we performed in order to check the implementation of the new infrastructure for the neutrino leakage scheme. 
Our reference physical system is a stable non-rotating NS (TOV).
In particular, we consider an NS with mass $1.68$\,M$_\odot$ and EOS LS220 \cite{LattimerSwesty1991}, which gives a radius of about $9.7$\,km.
The initial data are produced using the \texttt{Lorene/Nrotstar} code, as discussed in \Sref{idprod}. 
We consider both magnetized and non-magnetized NSs. For the latter, we initially add a purely poloidal magnetic field using the following vector potential prescription:
\begin{equation}
	\label{VecPot}
	A_\phi \equiv A_\mathrm{b} \varpi^2 {\rm max} \left( P - P_\mathrm{cut}, 0 \right)^{n_s} \ ,
\end{equation}
where $\varpi$ is the cylindrical radius, $A_\mathrm{b}$ is a positive constant, $P_\mathrm{cut}=0.04\, P_\mathrm{max}$ determines the cutoff when the magnetic field goes to zero inside the NS,
with $P_\mathrm{max}$ corresponding to the initial maximum gas pressure, and $n_s=2$ sets the degree of differentiability of the magnetic field strength~\cite{giacomazzo2011accurate}. The magnetic field is confined within the NS because of our use of the ideal MHD approximation, which is not valid in extremely low density regions (i.e~outside the NS). This is also the ``standard" magnetic field configuration used for the initial data of most BNS merger simulations (but see, e.g., \cite{Ruiz2016}).
The value of $A_b$ is chosen such that the maximum value of the initial magnetic field strength is set to $10^{16} \ \mathrm{G}$. This corresponds to the largest order of magnitude for a magnetic field that can be added to a TOV solution without introducing significant violations in the constraints of Einstein's equations. Significantly larger magnetic fields would indeed affect the structure of the star and therefore TOV equations could not be used anymore~\cite{Bocquet:1995je}. We also note that the average magnetic field that is reached in a post-merger remnant is typically of order $\sim10^{16} \ \mathrm{G}$ (see for example~\cite{Giacomazzo:2014qba}). One example of the initial and final magnetic field distribution is given in \Fref{FigBfield2D}.
\begin{table}[t!]
\caption{\label{table:IDB0}Initial Data used for the unmagnetised ($B=0$) tests.}
\begin{tabular}{@{}llccc}
\br
ID & Test Name & $\beta$-eq. Initial Data & $\nu$ Leakage & $T$ evolution \\
\mr
01 & \texttt{Spr\_S\_NL\_NB\_3D}	& S-slice 1$k_b$/bar	& Disabled	& Yes \\
02 & \texttt{GRH\_S\_NL\_NB}	& S-slice 1$k_b$/bar	& Disabled	& Yes \\ 
03 & \texttt{Spr\_S\_NL\_NB}	& S-slice 1$k_b$/bar	& Disabled	& Yes \\ 

04 & \texttt{Spr\_S\_YL\_NB\_3D}	& S-slice 1$k_b$/bar	& Enabled	& Yes \\ 
05 & \texttt{GRH\_S\_YL\_NB}	& S-slice 1$k_b$/bar	& Enabled	& Yes \\ 
06 & \texttt{Spr\_S\_YL\_NB}	& S-slice 1$k_b$/bar	& Enabled	& Yes \\ 

07 & \texttt{GRH\_T\_NL\_NB}	& T-slice $0.01$ MeV	& Disabled	& Yes \\ 
08 & \texttt{Spr\_T\_NL\_NB}	& T-slice $0.01$ MeV	& Disabled	& Yes \\ 

09 & \texttt{Spr\_T1\_NL\_NB}	& T-slice $0.01$ MeV	& Disabled	& Yes (after $t=2$ms) \\

10 & \texttt{GRH\_T\_YL\_NB}	& T-slice $0.01$ MeV	& Enabled	& Yes  \\ 
11 & \texttt{Spr\_T\_YL\_NB}	& T-slice $0.01$ MeV	& Enabled	& Yes  \\ 
12 & \texttt{Spr\_T1\_YL\_NB}	& T-slice $0.01$ MeV	& Enabled (at $t=3$ms) & Yes (after $t=2$ms) \\
\br
\end{tabular}\\
\end{table}
\begin{table}[h!]
	\caption{\label{table:IDB15}Initial Data used for the magnetized ($B\sim 10^{16}$G) tests.}
	\begin{tabular}{@{}llccc} 
		\br
		ID & Test Name & $\beta$-eq. Initial Data & $\nu$ Leakage & $T$ evolution \\ 
		\mr
		13 & \texttt{Spr\_S\_NL\_YB}	& S-slice 1$k_b$/bar	& Disabled	& Yes \\ 
		14 & \texttt{Spr\_S\_YL\_YB}	& S-slice 1$k_b$/bar	& Enabled	& Yes \\ 
		
		15 & \texttt{Spr\_T1\_NL\_YB}	& T-slice $0.01$ MeV	& Disabled	& Yes (after $t=2$ms) \\ 
		16 & \texttt{Spr\_T1\_YL\_YB}	& T-slice $0.01$ MeV	& Enabled (at $t=3$ms)	& Yes (after $t=2$ms) \\
		
		\br
	\end{tabular}\\
\end{table}

All the simulations adopt $5$ refinement levels. The outer boundary of the domain extends to $\approx193$\,km in every direction, while the innermost refinement level extends up to 13\,km. The finest grid resolution is $dx\approx177$\,m and the grid spacing doubles going from a refinement level to the next.
The entire NS is contained within the most refined region and the NS radius is covered with about 60 points.
Magnetized simulations adopt the full 3D domain,\footnote{
This choice is due to the lack of proper reflection symmetry conditions implemented for staggered variables (i.e.~for the vector and scalar potentials evolved by our code when magnetic fields are present).} 
while non-magnetized simulations are performed in octant symmetry, unless specified otherwise (label ``3D'' appearing in the test name). All simulations adopt the so called ``none" outer boundary conditions described in~\cite{cipolletta2020spritz} for the hydro variables (i.e., the values of all hydro variables are kept fixed to their initial values), linear extrapolation for the vector and scalar potential~\cite{cipolletta2020spritz},  and radiative boundary conditions for the metric variables~\cite{loffler2012einstein}. The simulations in octant symmetry also employ reflection symmetry conditions across the $x=0$, $y=0$, and $z=0$ planes.
For the ray-by-ray calculations of the neutrino leakage scheme, we use $9$ independent directions in $\theta$ and $16$ in $\phi$. While this holds for full 3D simulations, these numbers should be rescaled for cases where octant symmetry is employed (i.e.~$5$ independent directions in both $\theta$ and $\phi$).  

The set of tests we performed are summarized in Table \ref{table:IDB0} and \ref{table:IDB15}, referring to non-magnetized and magnetized cases, respectively.
All simulations cover about 6\,ms of evolution. This timescale corresponds to $\sim 14$ dynamical timescales and therefore it allows us to study these systems for a sufficiently long time for the tests presented here without requiring too much computational resources. We remark that the code was stopped after $\sim 6$ ms and it did not present any sign of instability or numerical problem at that time. In the following, we first discuss the results without neutrino leakage, testing the implementation of the tabulated EOS handling, and then those with neutrino leakage, with and without neutrino heating. 

Among the physical quantities monitored in our tests, we considered the total neutrino luminosity of each neutrino species, defined in cartesian coordinates as 
\begin{equation}
\label{Linfcart}
\eqalign{
L_{\nu_i}^\infty = \int^{\infty}_0 \int^{\infty}_0 \int^{\infty}_0 & Q^{\rm{eff}}_{\nu_i} (x',y',z') \left[ \alpha^2(x',y',z') W(x',y',z') \right. \cr  
& \left. \times\left( 1 + v^r(x',y',z') \right) \right] \sqrt{\gamma} dx'dy'dz',
}
\end{equation}
where $v^r = (xv^x + yv^y  + zv^z) /\sqrt{x^2 + y^2 + z^2}$.

\subsection{Testing Tabulated EOS Without Neutrino Leakage}

In order to test the implementation of the tabulated EOS treatment, we here report the results of all the simulations performed without enabling the leakage scheme, starting from both S--slicing and T--slicing initial data.

\begin{figure}[tbh!]
\begin{center}
\includegraphics[width=0.95\linewidth]{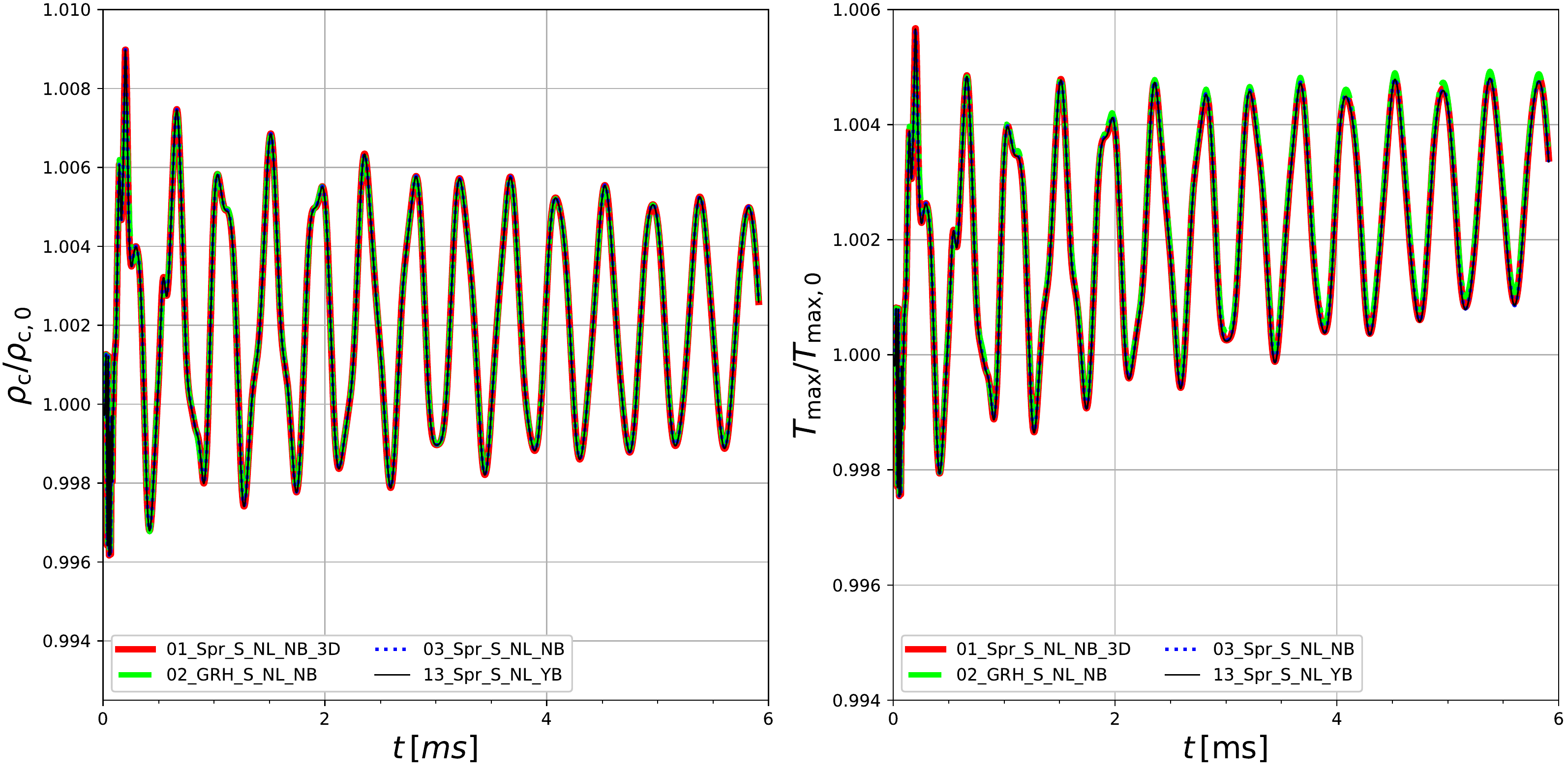}
\caption{\label{FigS}Evolution of initial data produced with the S--slicing conditions and without neutrinos. The left panel shows the evolution of the maximum rest-mass density normalized to its initial value. The right panel is the equivalent for the maximum temperature (which is located at the NS centre).}
\end{center}
\end{figure}

The results for the evolution of the maximum of $\rho$ and $T$ for the S--slicing initial condition are shown in \Fref{FigS}. In these models the maximum of the temperature is located at the NS centre and it shows an increase of less than 1\% by the end of the simulation (likely due to shocks produced by the NS oscillations). In particular, the figure shows exact match for simulations 01, 02, 03, and 13, as expected (see Table~\ref{table:IDB0} and \ref{table:IDB15}). Noticeably, adopting octant symmetry in pure-hydro simulations 02 and 03, performed with the \texttt{GRHydro} and the \texttt{Spritz} codes respectively, produces the same results as adopting full-3D in simulations 01 and 13. Moreover, the magnetic field of simulation 13 is correctly handled during the evolution and does not significantly alter the hydrodynamic quantities as expected (we remind that, even if large, a magnetic field of $\sim 10^{16}$\,G provides a magnetic energy which is still $\sim 2$ orders of magnitude below equipartition). 

\begin{figure}[tbh!]
\begin{center}
\includegraphics[width=0.95\linewidth]{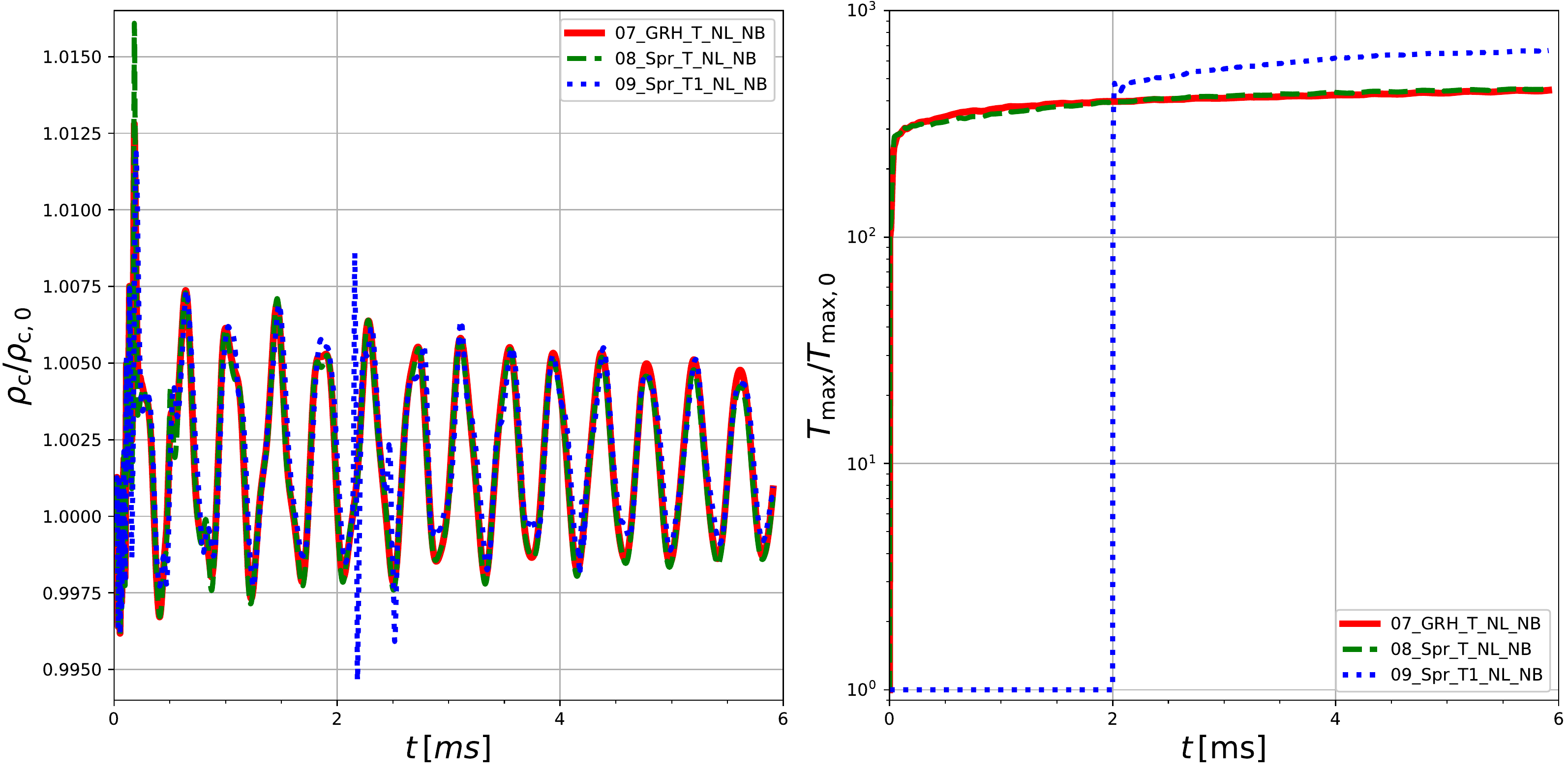}
\caption{\label{FigT}Same as \Fref{FigS} but for T--slicing conditions (in this case the maximum of the temperature is located on the NS surface).}
\end{center}
\end{figure}

\begin{figure}[tbh!]
\begin{center}
\includegraphics[width=0.95\linewidth]{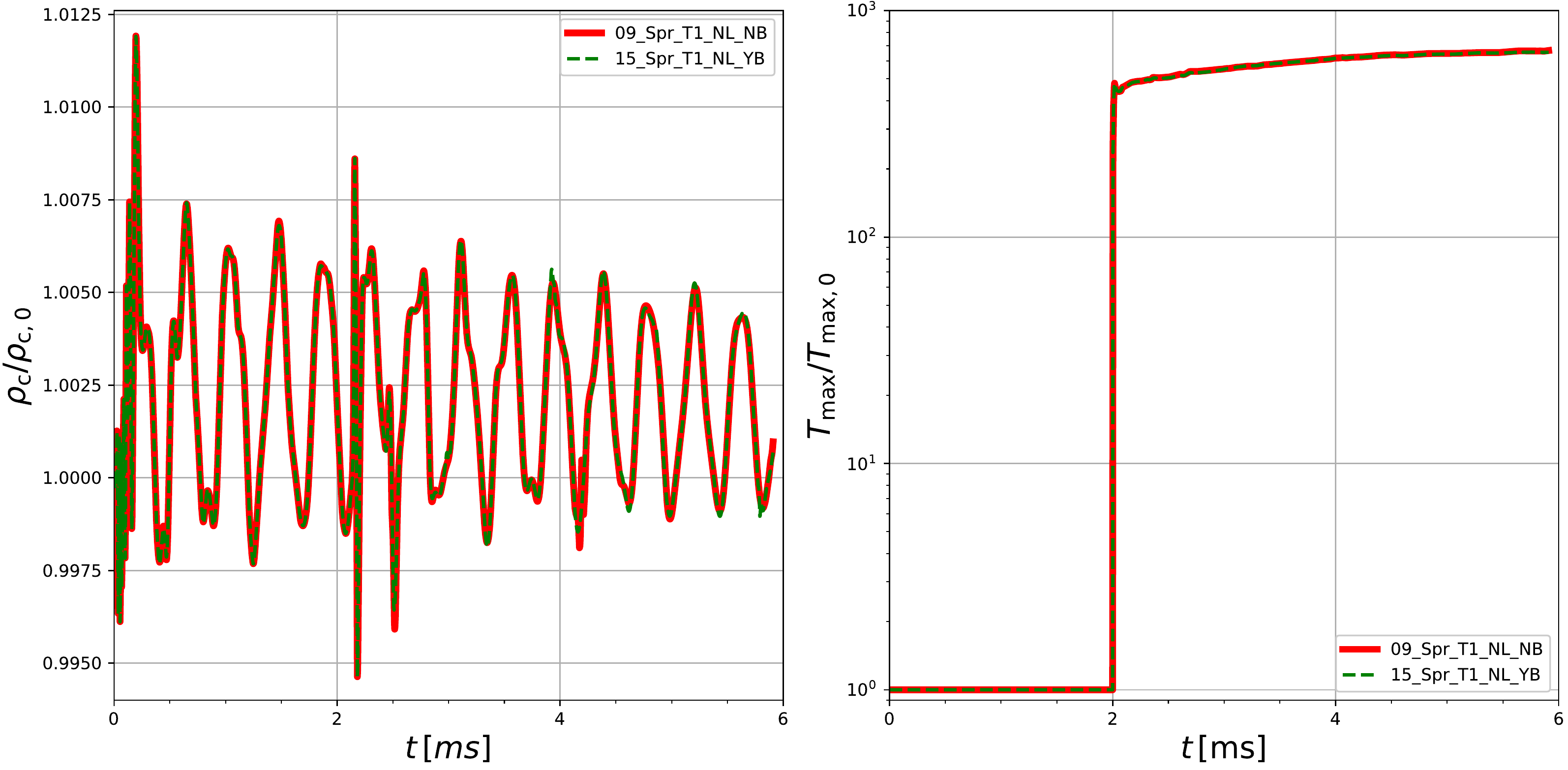}
\caption{\label{FigSTomp}Evolution of initial data produced with the T--slicing conditions for simulations where the temperature evolution is only allowed after $2$\,ms. The left panel shows the evolution of the maximum rest-mass density normalized to its initial value, while the right panel shows the same but for the maximum temperature.}
\end{center}
\end{figure}

The same comparison for T--slicing initial condition is shown in \Fref{FigT} and \Fref{FigSTomp}. In this case the maximum of the temperature is located instead on the NS surface. Simulation 09 is the most delicate in the pure-hydro setting, since it forces the temperature $T$ to be constant for the first $\sim 2$\,ms and then allows it to evolve (see~\Sref{C2P}). When the temperature is free to evolve, an artificial shock is produced at the surface of the NS (as expected), but, after this initial transient, the maximum of $\rho$ follows closely the results given by the simulations 07 and 08, where $T$ is evolved since the beginning. Also the temperature, after the initial transient, tends to a constant value. In addition, \Fref{FigSTomp} shows perfect match between simulation 09, performed in pure-hydro, and the magnetized simulation 15.

Based on the above results, we conclude that the tabulated EOS treatment is correctly handled by our implementation and we can then proceed in testing the neutrino leakage scheme.

\subsection{Testing the Neutrino Leakage Implementation}

Here we report the results of simulations involving neutrino leakage with constant-$S$ and constant-$T$ initial data, including the evolution of the total neutrino luminosity for each neutrino species, computed according to \Eref{Linfcart}. We first present the results of tests performed without the heating contribution of \Eref{qheat} and then including it.

\subsubsection{Tests Without Heating.}

\begin{figure}[tbh!]
\begin{center}
\includegraphics[width=0.95\linewidth]{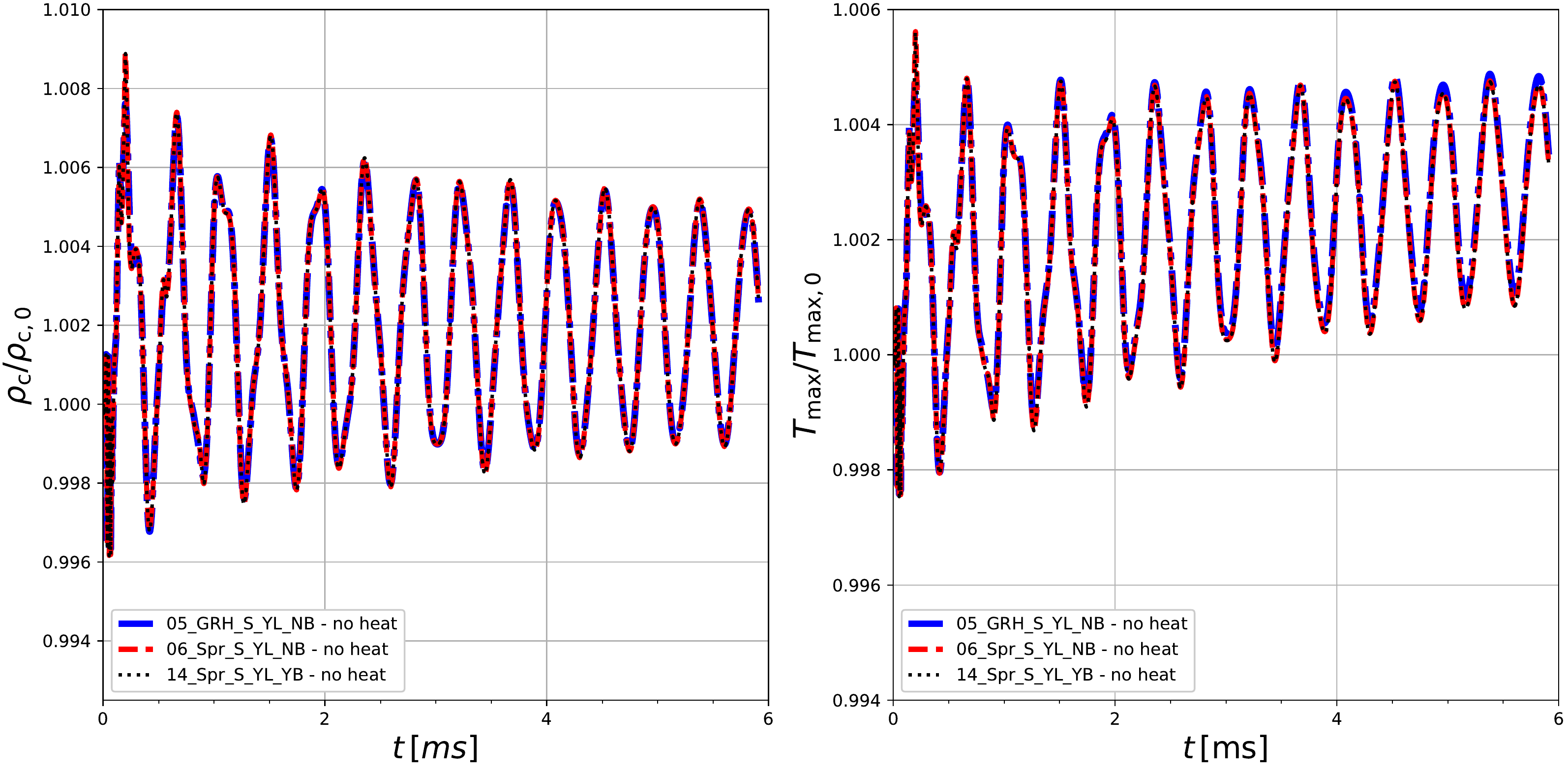}
\includegraphics[width=0.92\linewidth]{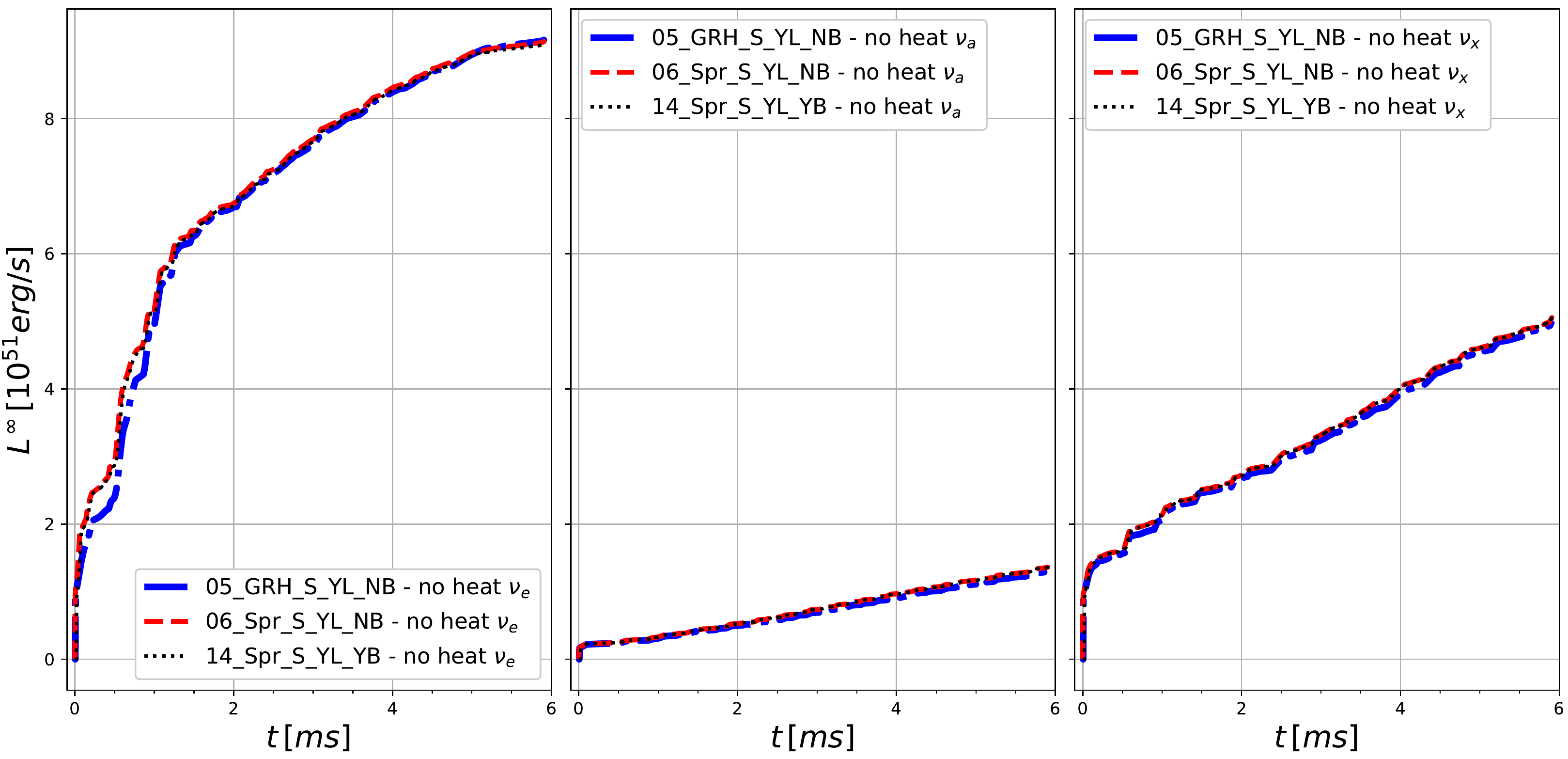}
\caption{\label{FigSleakNoHeat}\textit{Top panels:} Same as \Fref{FigS} but for $S$--slicing simulations considering leakage and no heating. \textit{Bottom panels:} Evolution of neutrino luminosities (\Eref{Linfcart}) for the different neutrino species (from left to right: $\nu_e$, $\bar{\nu}_e$ indicated here as $\nu_a$, and $\nu_x$).}
\end{center}
\end{figure}

\Fref{FigSleakNoHeat} shows the comparison of tests evolving $S$--slicing initial data with neutrino leakage, but without the contribution of neutrino absorption and heating: the maxima of $\rho$ and $T$ normalized to their initial values are shown in the top panels, while the bottom panels show the results for the luminosity of each neutrino species (electron neutrinos, electron antineutrinos, and the $\mu$ and $\tau$ species going from left to right) as computed in \Eref{Linfcart}. In particular, the luminosity plots show that the scenario is clearly dominated by electron capture. Also in this case we can see that the maximum temperature, which for the $S$--slicing initial data is located at the NS centre, shows an increase of less than 1\%. Neutrino cooling at the centre of the star is not effective due to the high density (and thus high optical depths) and therefore it does not significantly affect the temperature evolution in that region.

\begin{figure}[tbh!]
\begin{center}
\includegraphics[width=0.95\linewidth]{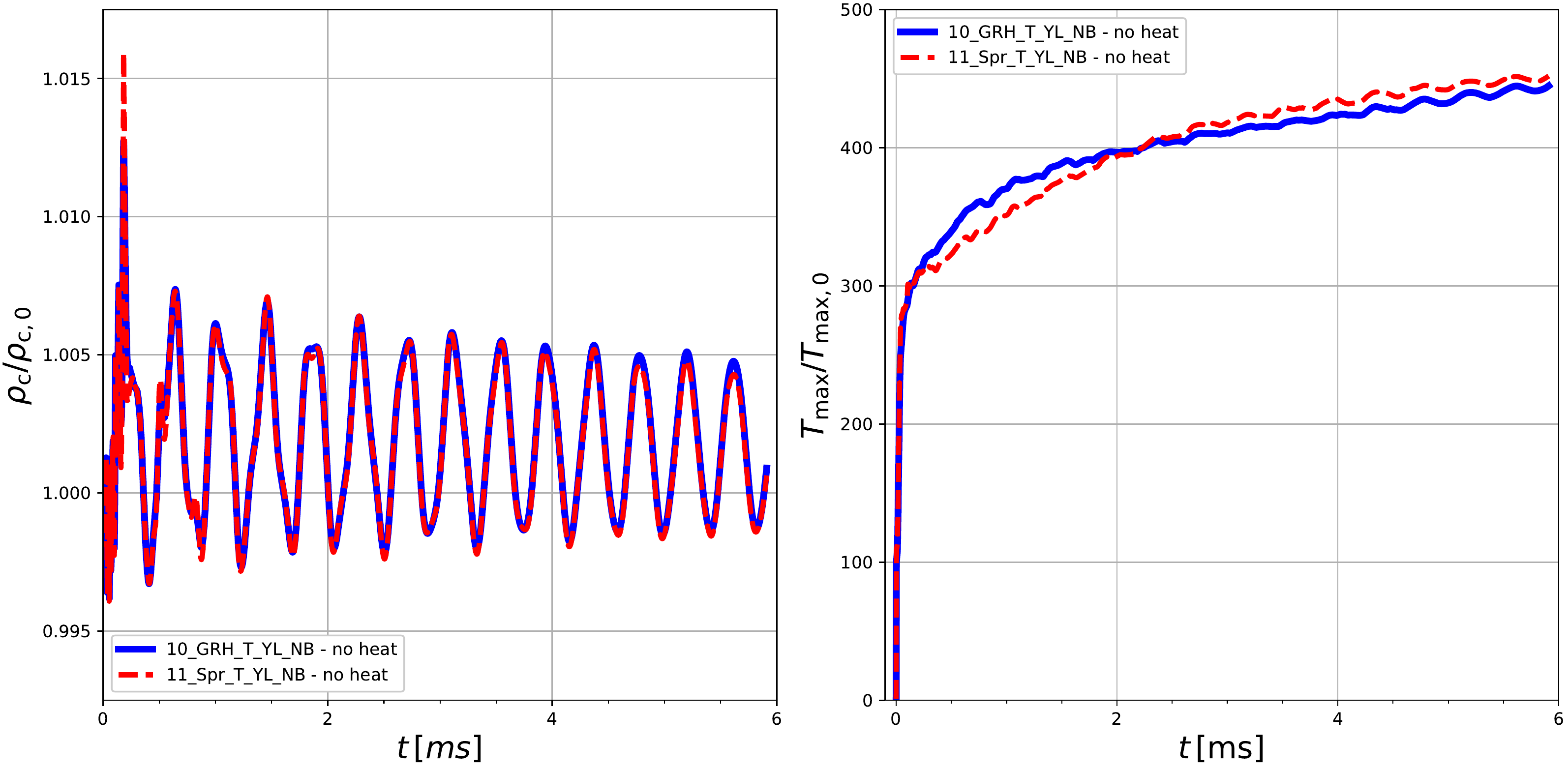}
\includegraphics[width=0.92\linewidth]{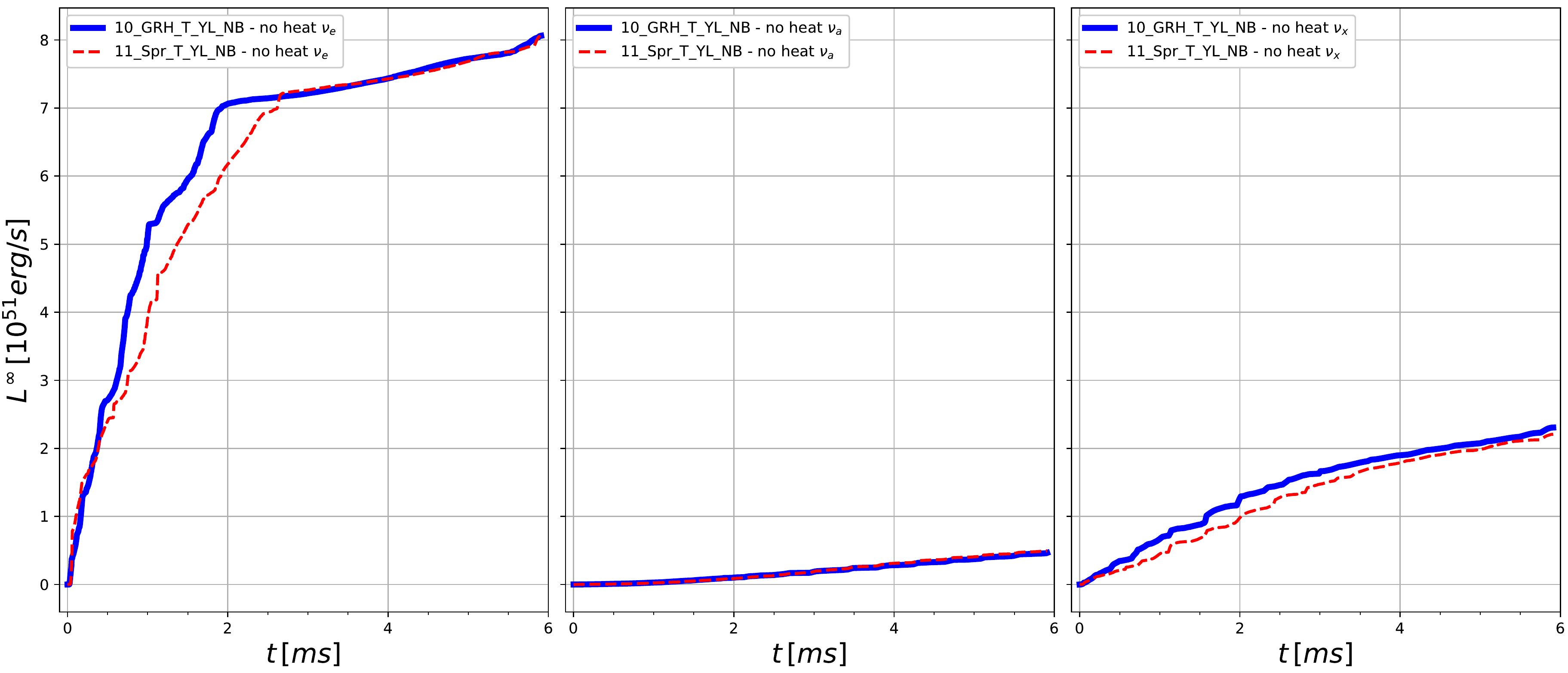}
\caption{\label{FigTleakNoHeat}Same as \Fref{FigSleakNoHeat} but for T--slicing simulations 10 and 11, performed respectively with the \texttt{GRHydro} and the \texttt{Spritz} codes.}
\end{center}
\end{figure}

A similar comparison for $T$--slicing initial data is shown in \Fref{FigTleakNoHeat} (we remind that in this case the maximum of the temperature is located on the NS surface and it is strongly affected by the artificial shocks that develop there). Despite minor differences due to the different implementations in the \texttt{GRHydro} and \texttt{Spritz} codes, the results appear in good agreement.

\subsubsection{Tests Including Heating.}

\begin{figure}[tbh!]
\begin{center}
\includegraphics[width=0.95\linewidth]{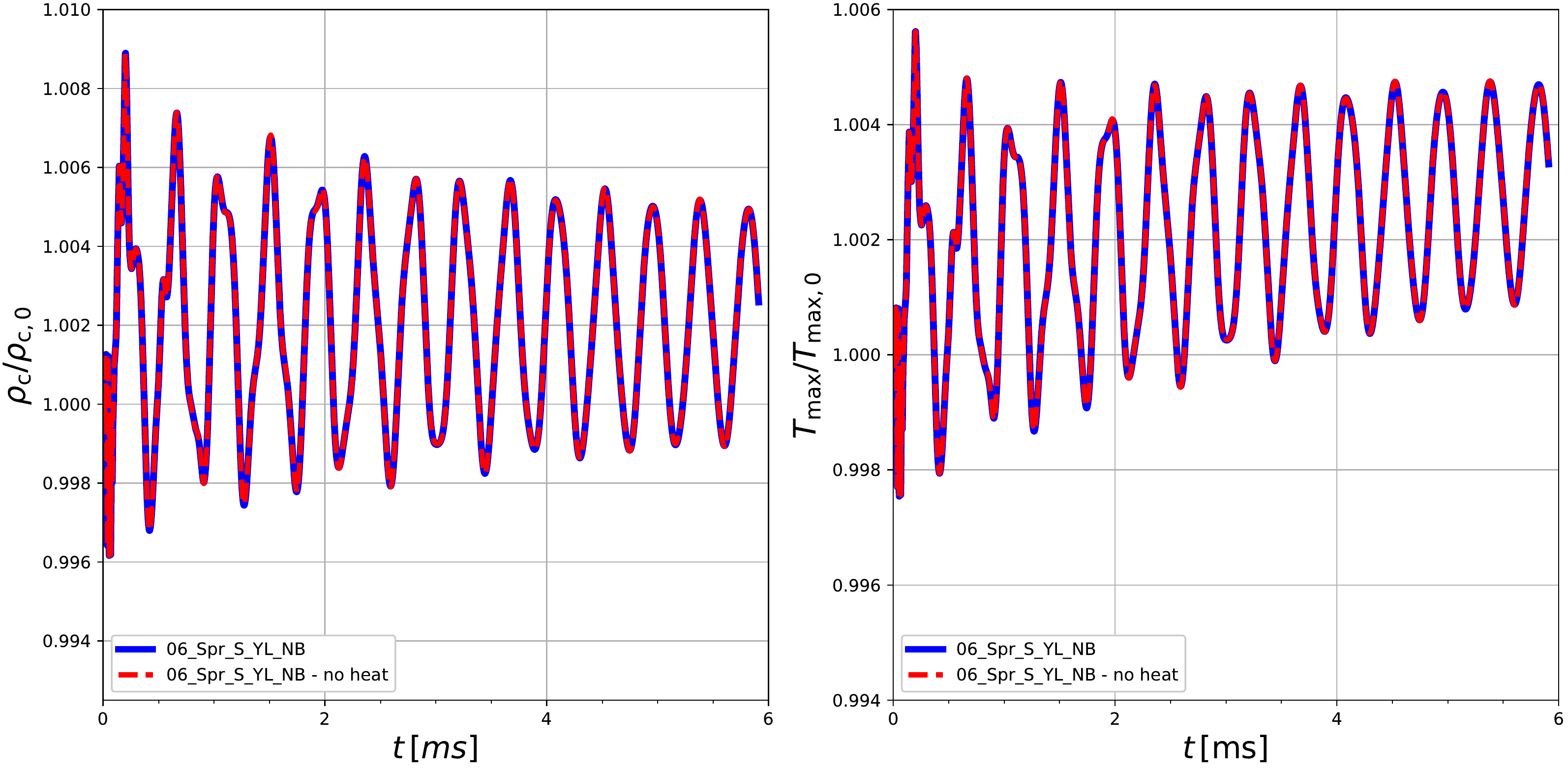}
\includegraphics[width=0.92\linewidth]{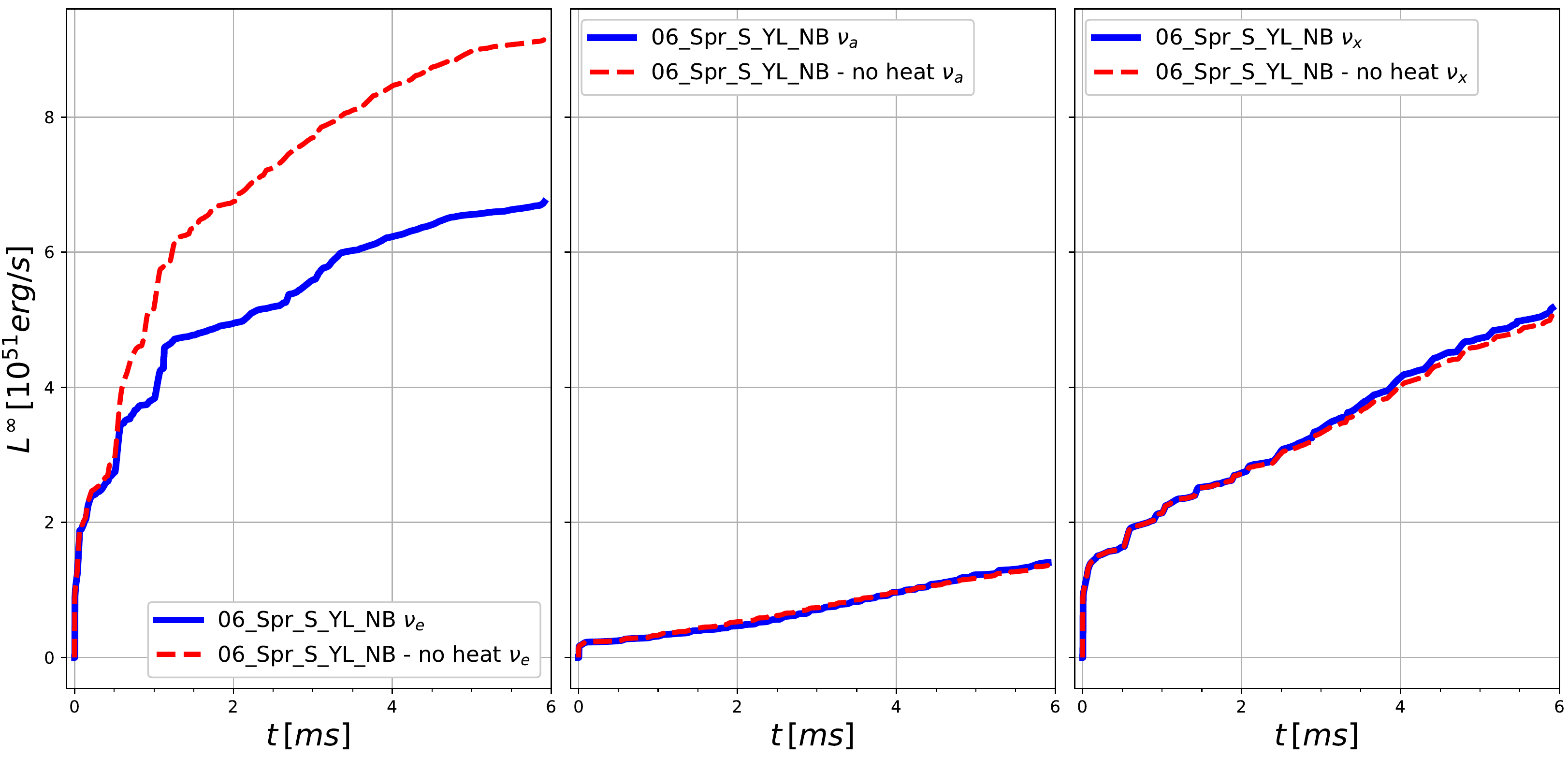}
\caption{\label{Sheatvsnoheat}Same as \Fref{FigSleakNoHeat} but for the simulation 06 including leakage, with and without the heating contribution.}
\end{center}
\end{figure}

\begin{figure}[tbh!]
\begin{center}
\includegraphics[width=0.95\linewidth]{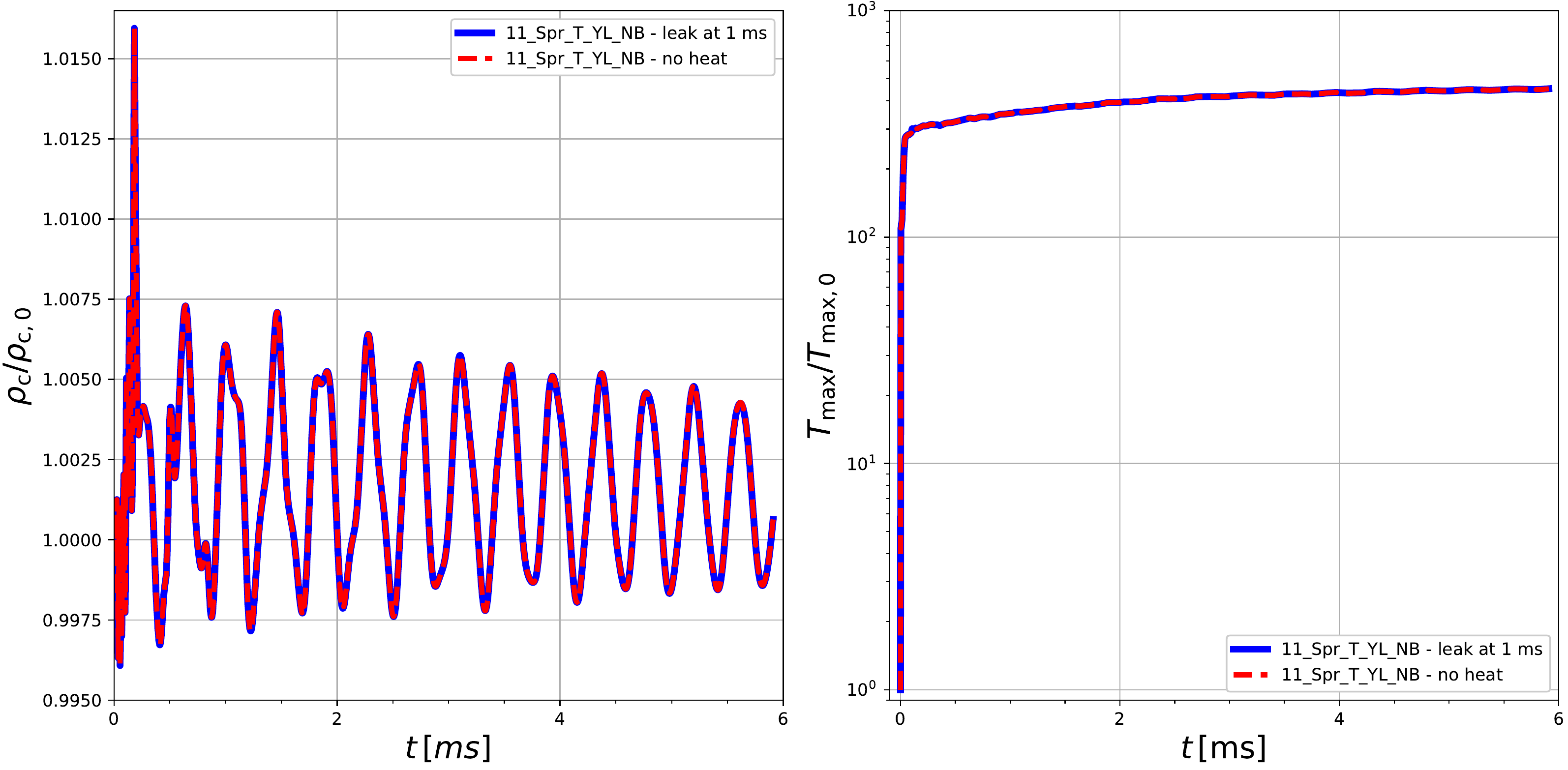}
\includegraphics[width=0.92\linewidth]{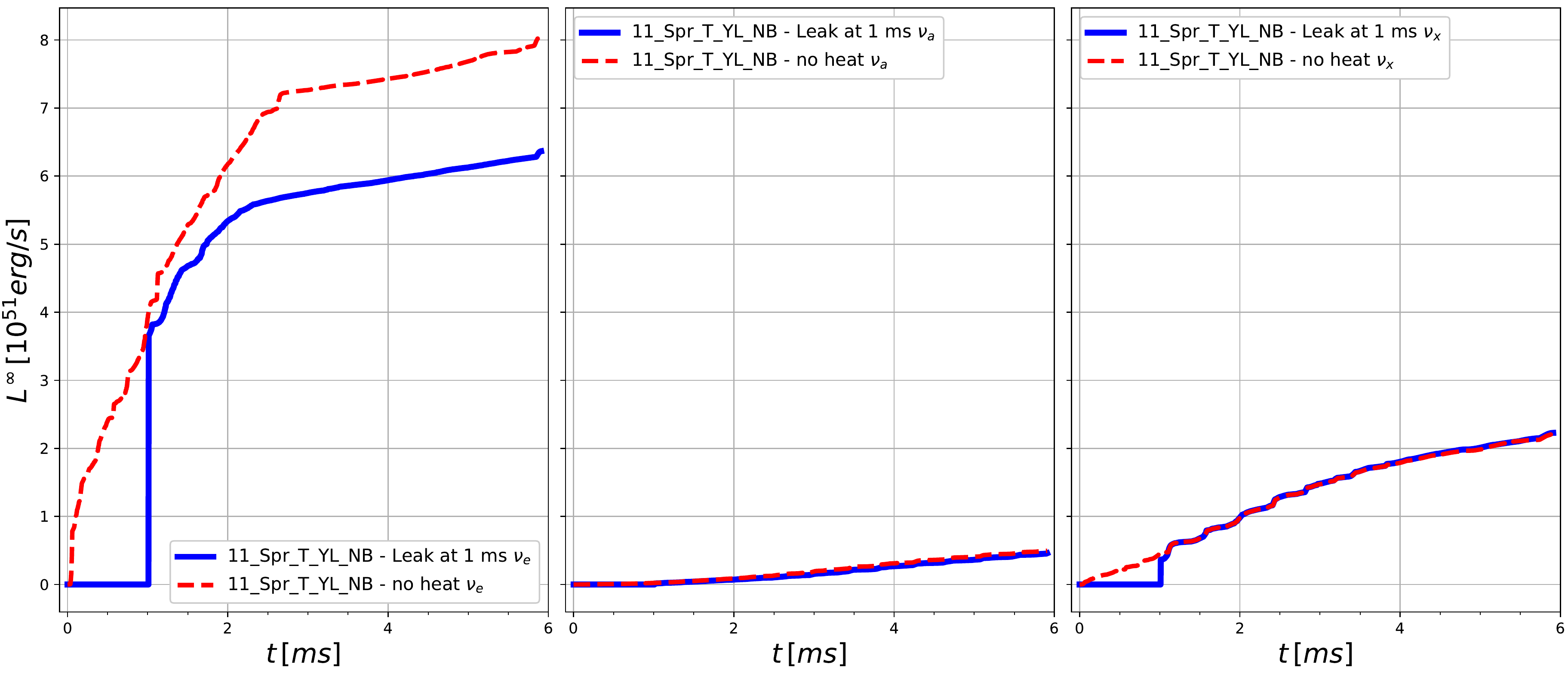}
\caption{\label{Theatvsnoheat}Same as \Fref{FigSleakNoHeat} but for the simulation 10 including leakage, with and without the heating contribution. Where heating is considered (blue solid curve), the leakage is activated at $t = 1$\,ms in order to avoid spurious effects due to the initial sharp drift in the maximum temperature.}
\end{center}
\end{figure}

We now turn to consider how the heating contribution alters the results of simulations. In \Fref{Sheatvsnoheat} and \Fref{Theatvsnoheat} we compare the results respectively of one $S$--slicing and one $T$--slicing ID performed with and without such contribution. As already seen in \Fref{FigT}, starting from cold NS initial data produces a sharp transient for the maximum of $T$ (located at the NS surface for the $T$--slicing ID) in the first few time steps, where the NS internal temperature undergoes a re-adjustment (due also to the expected production of shocks at the NS surface). This transition may be an issue when considering neutrino leakage since it may produce luminosities much larger than expected. Moreover, we recall that the heating given by \Eref{qheat} is not self-consistent in terms of energy balance (see also \Sref{nuimpl}).
Therefore, when considering the heating contribution (\Fref{Theatvsnoheat}), we activated the leakage $1$\,ms later, i.e.~after the initial transient.

\begin{figure}[tbh!]
\begin{center}
\includegraphics[width=0.95\linewidth]{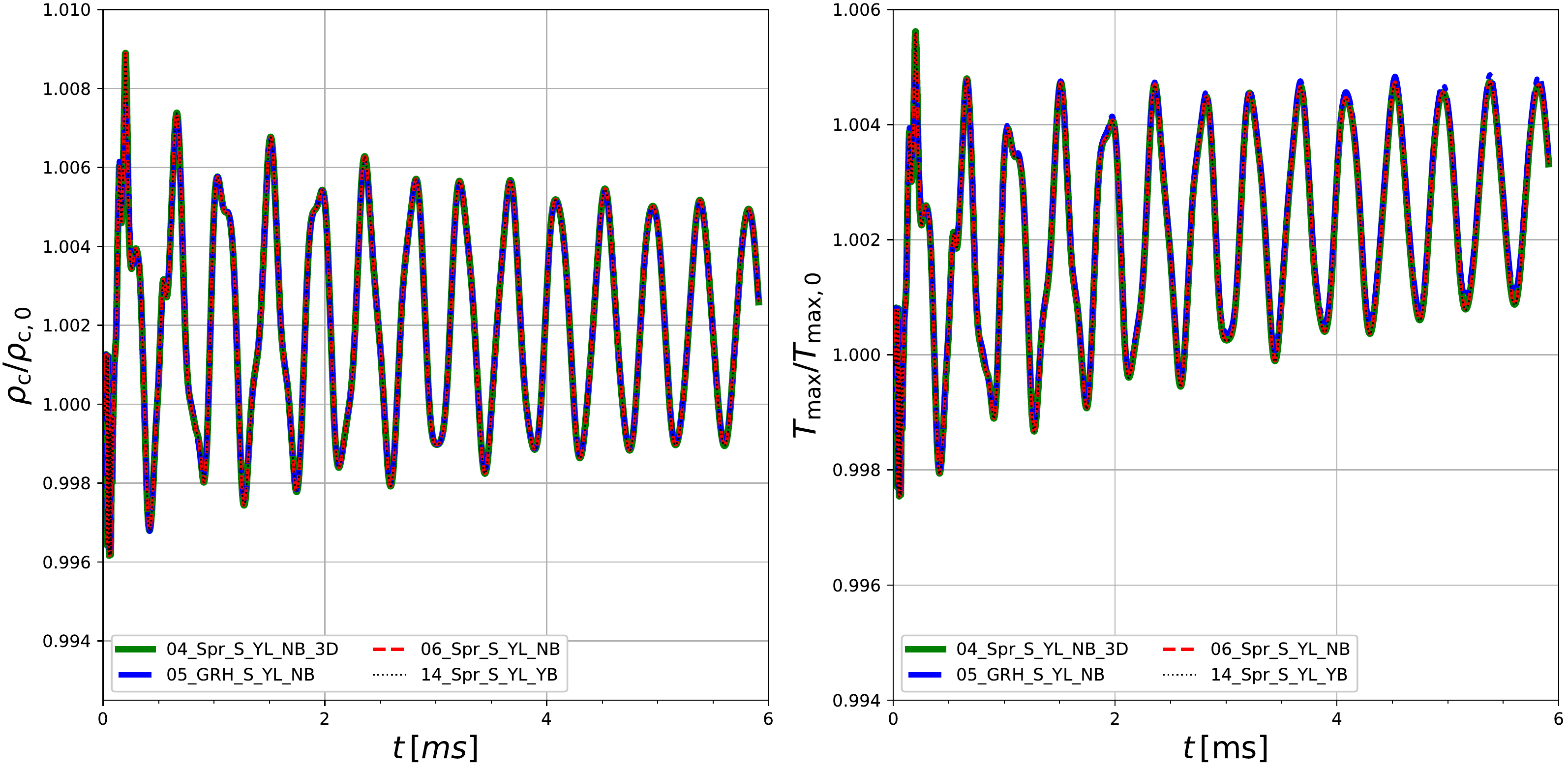}
\includegraphics[width=0.92\linewidth]{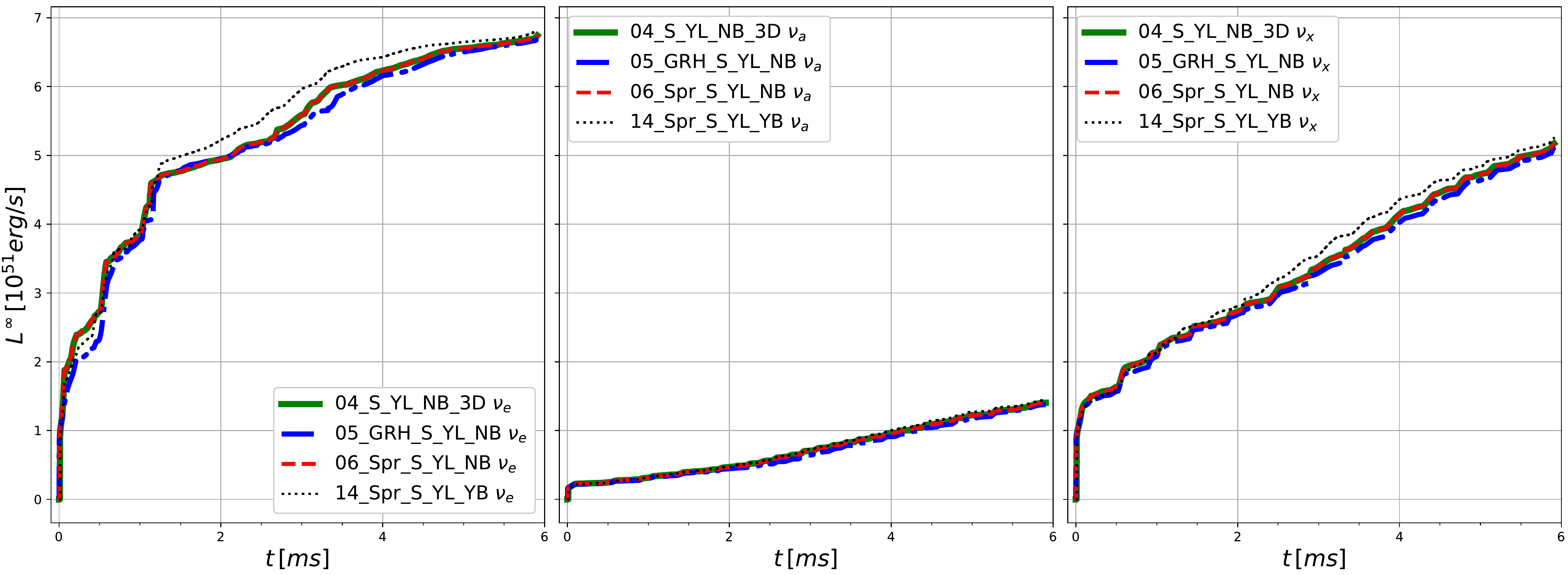}
\caption{\label{FigSleak}Same as \Fref{FigSleakNoHeat} but considering the heating contribution.}
\end{center}
\end{figure}

\begin{figure}[tbh!]
\begin{center}
\includegraphics[width=0.6\linewidth]{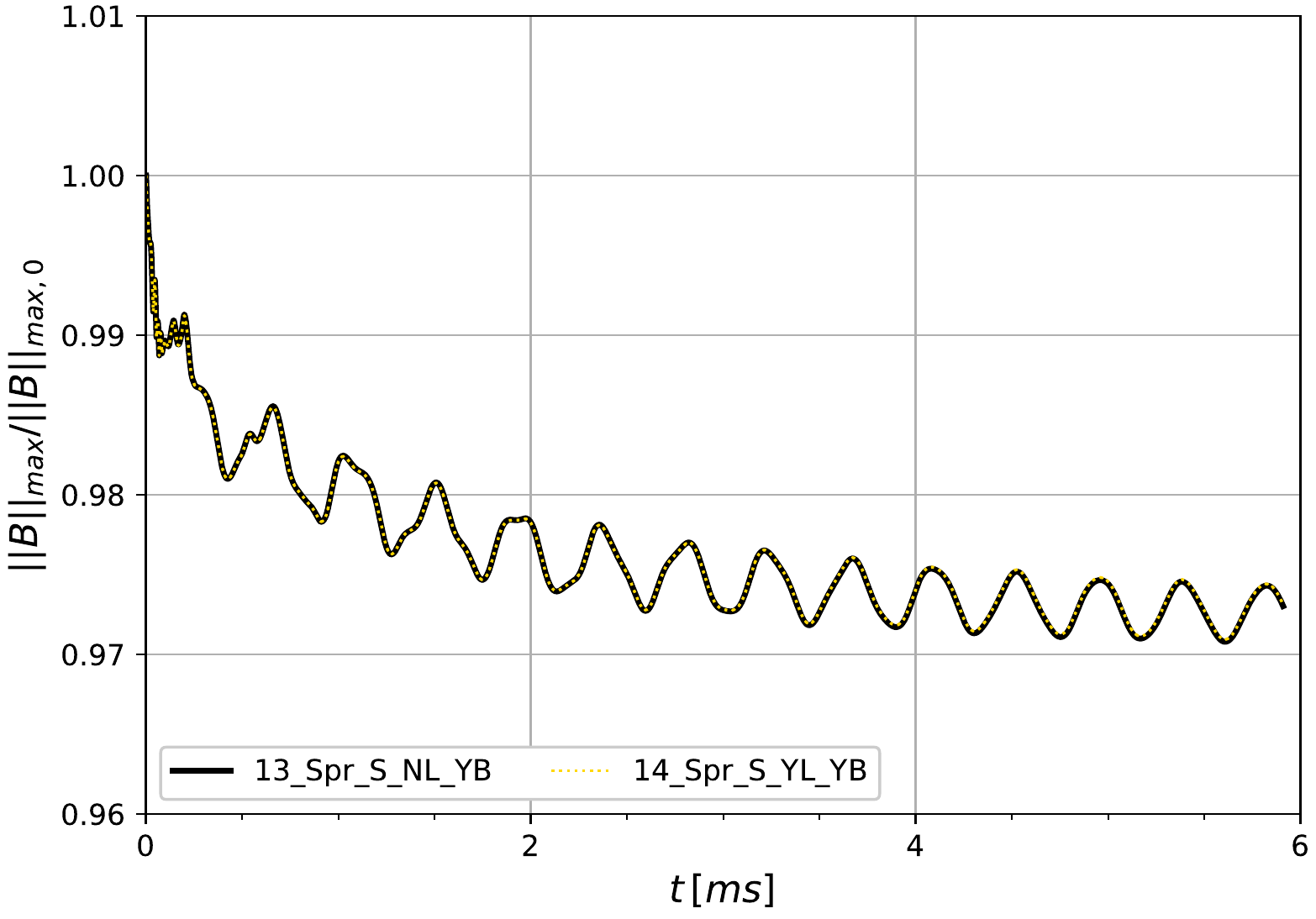}
\caption{\label{FigSBmax}Comparison of results for evolution of $B_{\rm{max}}$ produced with the S--slicing conditions, with and without neutrino leakage and heating.}
\end{center}
\end{figure}

\Fref{FigSleak} collect results for $S$--slicing ID and neutrino leakage including heating. In this case, without an initial temperature readjustment, the heating contribution does not need to be activated after 1\,ms.
We also show the maximum magnetic field evolution for the magnetized cases 13 (without leakage) and 14 (with leakage and the heating contribution) in \Fref{FigSBmax}. We found an exact match.

\begin{figure}[tbh!]
\begin{center}
\includegraphics[width=0.95\linewidth]{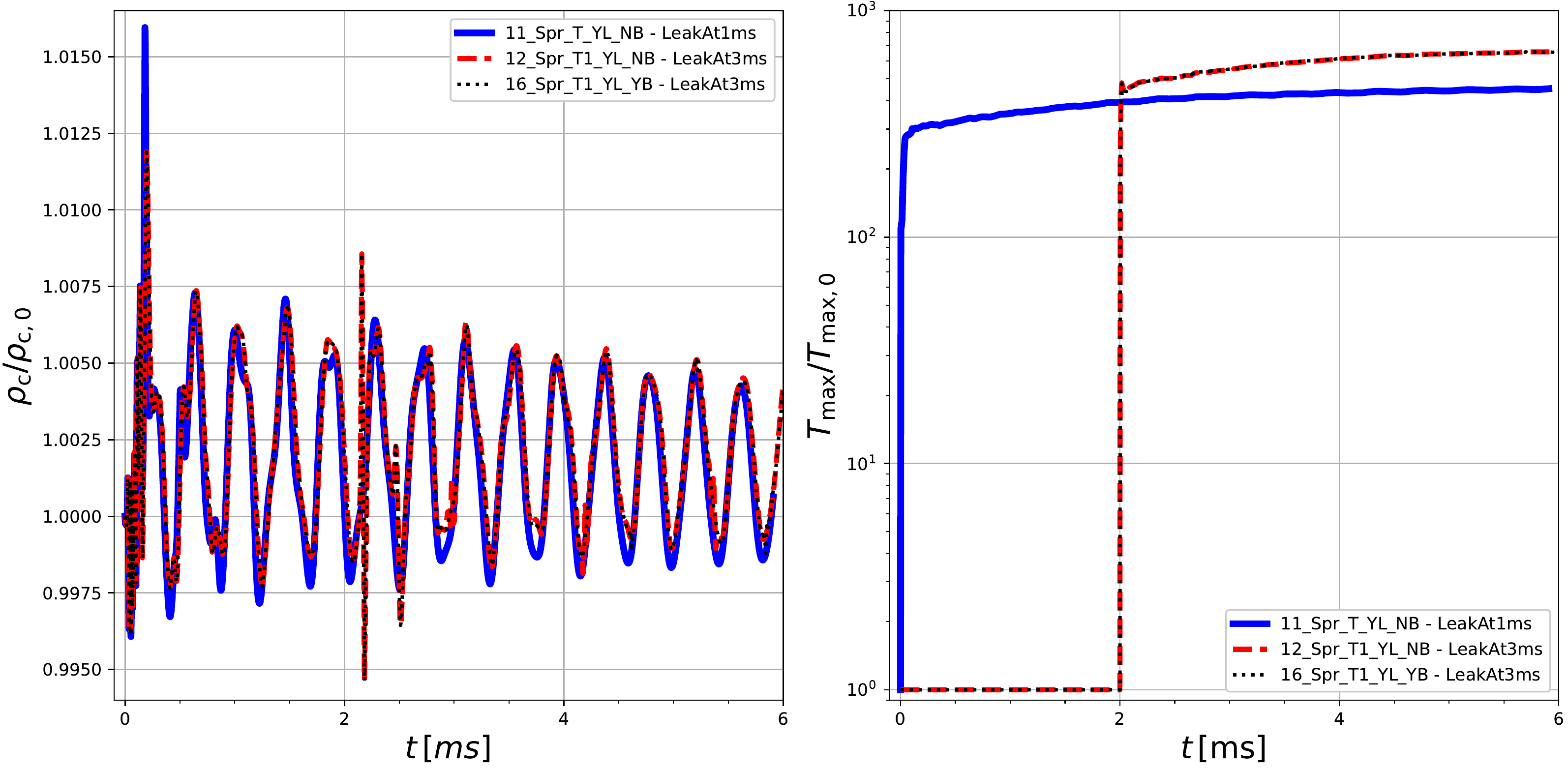}
\includegraphics[width=0.92\linewidth]{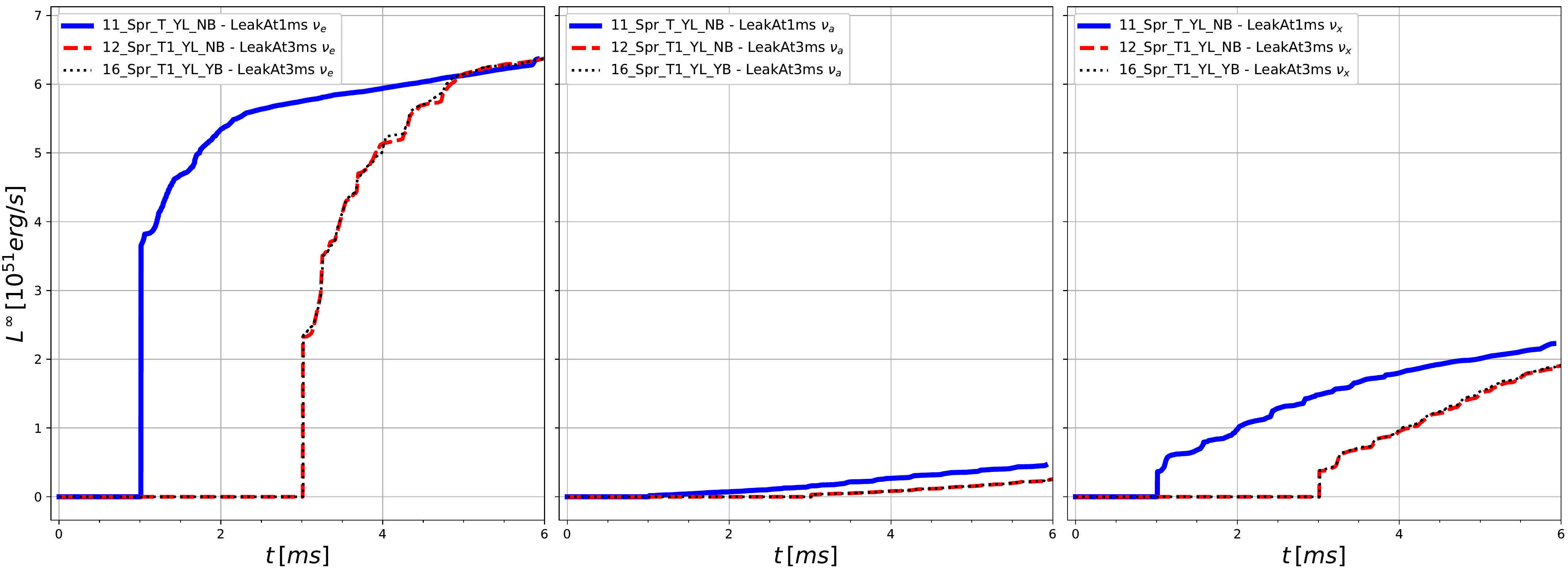}
\caption{\label{FigTleak}Same as \Fref{FigSleakNoHeat} but considering T-slicing cases with neutrino leakage and heating contribution, where leakage is activated 1\,ms after temperature evolution is enabled (at $t=0$ for model 11, at $t=2$\,ms for model 12, 16).}
\end{center}
\end{figure}

\begin{figure}[tbh!]
\begin{center}
\includegraphics[width=0.6\linewidth]{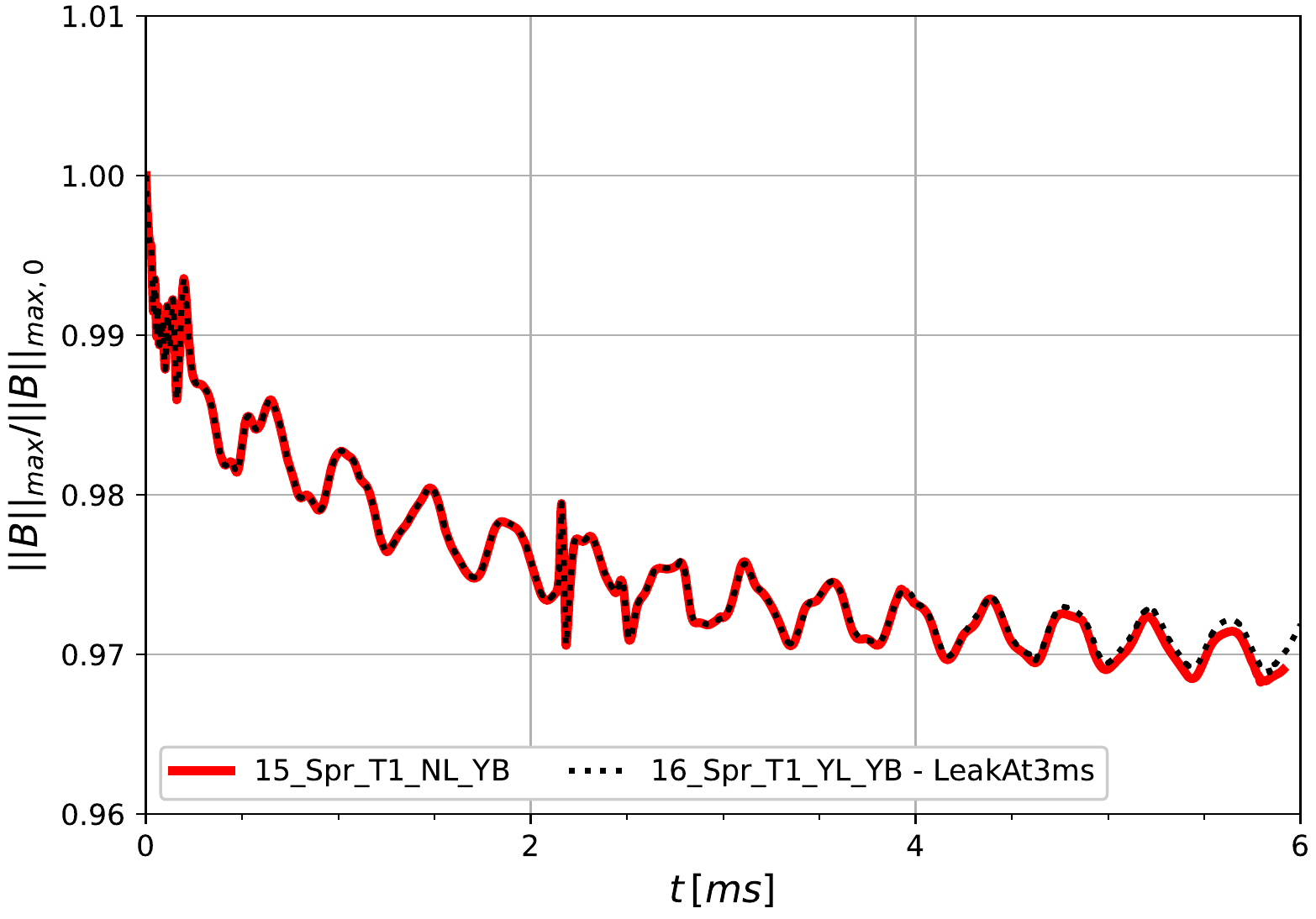}
\caption{\label{FigTBmax}Same as \Fref{FigSBmax} but for T--slicing ID.}
\end{center}
\end{figure}

In \Fref{FigTleak}, we compare the cases with cold NS initial data ($T$--slicing) and neutrino leakage including heating. For simulation 11, which evolves the temperature since the beginning, we enable the leakage after only $1$\,ms. For simulations 12 and 16, evolving the temperature only after 2\,ms, we enable the leakage at $3$\,ms. 
Despite the difference in the activation times of T evolution and leakage, and in the presence or absence of magnetic fields, all the results show a very good agreement in the maximum rest-mass density and the late-time electron neutrino luminosities.
Finally, looking again at the maximum magnetic field evolution, \Fref{FigTBmax} shows that also simulations 15 and 16 are perfectly matching each other.

All the test results presented in this Section are indicative of a correct implementation of the neutrino leakage scheme and that the code is ready to be used in more complex astrophysical scenarios, e.g., BNS mergers including tabulated EOS, magnetic fields, and neutrino emission and absorption (with the intrinsic limitations of the leakage scheme itself; see discussion below).

\clearpage

\section{Discussion and Conclusions\label{conclusions}}

We presented a new version of our fully GRMHD code \texttt{Spritz} (available on Zenodo as version 1.1.0~\cite{spritz}) that now includes neutrino cooling and heating via the \texttt{ZelmaniLeak} code. We performed a series of tests to show the robustness of the code in handling a variety of different physical scenarios, including the evolution of both ``cold" and ``hot" NSs with and without magnetic fields or neutrino leakage. For the cases with neutrino leakage, we also considered the effects of having neutrino heating activated or deactivated.

The \texttt{Spritz} code will be used in future work to study the merger of magnetized BNS systems employing finite temperature tabulated EOSs and including neutrino emission and reabsorption. The code has indeed all the necessary routines to evolve BNS systems during inspiral, merger and post-merger phases. Initial data for BNS systems can be produced with the publicly available \texttt{LORENE} library and they can be read with the \texttt{EinsteinInitialData/Meudon\_Bin\_NS} thorn included in the \texttt{Einstein Toolkit}. Results from BNS merger simulations with \texttt{Spritz} will be presented in a future paper.
We note that the neutrino leakage scheme implemented here, which represents the first step towards a more advanced neutrino treatment, presents some limitations. 
First, the method adopts a ray-by-ray approach, which is well-suited for problems involving geometries that are, at first approximation, spherically symmetric (for instance, in the context of core collapse supernovae; see, e.g., \cite{ott2012correlated} and references therein). 
For this reason, it should work reasonably well in a post-merger remnant NS phase where the latter has already achieved an approximately spherical configuration \cite{Moesta2020}, but in the early post-merger or after the collapse into a BH surrounded by an accretion disk, when significant deviations from spherical symmetry are present, it would in part over-estimate the neutrino opacities used in the leakage scheme. 
To overcome such limitation, various groups implemented a local opacity calculation \cite{Neilsen2014}, which better accounts for non-spherical geometries. 
This different opacity calculation has been already employed in magnetized BNS mergers with neutrino leakage \cite{palenzuela2015effects,Most2019}, but without accounting for neutrino heating/reabsorption. These simulations represent the current state-of-the-art in the context of magnetized BNS mergers with neutrinos. 
A second and more general limitation, that is shared among all leakage schemes, is that neutrino energy estimates are not precise enough to provide an accurate estimate of the electron fraction in the ejecta and thus in the computation of the r-process nucleosynthesis and consequent kilonova emission (e.g., see~\cite{Foucart:2016rxm}).
The above limitations can be overcome by adopting more accurate neutrino transport schemes, such as the Monte-Carlo-based scheme recently adopted for the first time in (nonmagnetized) BNS merger simulations~\cite{Foucart2020} or even the (much more computationally expensive) full solution of Boltzmann transport equations~\cite{Shibata:2014vza}. 
Future work will be devoted to improve on our current neutrino treatment, possibly following the direction suggested by \cite{Foucart2020}.

We have also implemented high-order methods for the evolution of hydrodynamical quantities (see~\ref{highord} for a discussion) which will allow our code to provide a better description of matter dynamics and produce also more accurate GW signals. We plan to extend the implementation of these methods also to the equations describing the evolution of magnetic fields, following an approach similar to the one discussed in~\cite{Most2019}.

The initial data and EOSs used in this paper are available for download in the supplemental material.

\section{Acknowledgments}
We thank Ernazar Abdikamalov, Elias Most, Jerome Novak, Carlos Palenzuela, and Albino Perego for the very useful discussions. We also thank the two anonymous referees for their useful comments.
F.C. is funded through the NASA TCAN 80NSSC18K1488 grant.
J.V.K. kindly acknowledges the CARIPARO Foundation for funding his PhD fellowship within the PhD School in Physics at the University of Padova.
All the simulations were performed on GALILEO and MARCONI machines at CINECA. Some of the numerical calculations have been made possible through a CINECA-INFN agreement, providing the allocation \texttt{INF20\_teongrav}. Other simulations were performed with the following authors' allocations: \texttt{IsC77\_SPRITZ}, \texttt{IsB18\_BlueKN}, and \texttt{IsB21\_SPRITZ}.

\appendix

\section{Higher order methods}
\label{highord}

Here we present the implementation of the high-order scheme in the \texttt{Spritz} code and some tests that assess the convergence order of this algorithm.

\subsection{Reconstruction step: WENOZ method}
The first step in the development of a high-order scheme is the choice of the reconstruction method. Here, we consider the fifth-order WENOZ algorithm \cite{borges2008wenoz}. In the following, we will consider only one dimension without loss of generality: the multidimensional scheme is simply retrieved by considering the fluxes in each direction separately.

The fifth-order WENO scheme employs a 5-points stencil, $S^5$, which is subdivided into three 3-points substencils, $\{S^0, S^1, S^2\}$. 
The polynomial approximation $f_{i+1/2}$, which is the reconstruction of the grid function $f_i$ on the left side of the interface\footnote{$f_{i-1/2}$ is simply given by swapping the indices of the stencil: $(i-2,i-1,i,i+1,i+2) \rightarrow (i+2,i+1,i,i-1,i-2)$}, is built through the following convex combination of the interpolated values $f^k_{i+1/2}$, that are third degree polynomials defined on each substencil $S^k$, $k=0,1,2$:
\begin{equation}
f_{i+1/2} = \sum_{k=0}^2 \omega_k f^k_{i+1/2} \, .
\end{equation}
The polynomial on each substencil is given by the quadratic interpolations
\begin{equation}\label{f0}
f^0_{i+1/2} = \frac{1}{8}\left(3f_{i-2} - 10f_{i-1} + 15f_i\right)  \, ,
\end{equation}

\begin{equation}\label{f1}
f^1_{i+1/2} = \frac{1}{8}\left(-f_{i-1} + 6f_i + 3f_{i+1}\right)  \, ,
\end{equation}

\begin{equation}\label{f2}
f^2_{i+1/2} = \frac{1}{8}\left(3f_i + 6f_{i+1} - f_{i+2}\right)  \, .
\end{equation}
The weights $\omega_k$ are defined as
\begin{equation}
\omega_k = \frac{\alpha_k}{\sum_{j=0}^2 \alpha_j}  \, .
\end{equation}
For WENOZ, the unnormalized weights $\alpha_k$ are defined as
\begin{equation}
\alpha_k = d_k \left(1 + \frac{|\beta_0 - \beta_2|}{\beta_k + \varepsilon}\right) \, ,
\end{equation}
with $\varepsilon = 10^{-26}$ (which avoids a possible division by zero), optimal weights $d_k = (1/16, 10/16, 5/16)$, corresponding to the weights obtained for smooth fields, and smoothness indicators
\begin{equation}\label{beta0}
	\beta_0 = \frac{13}{12}\left(f_{i-2} - 2f_{i-1} + f_i\right)^2 + \frac{1}{4}\left(f_{i-2} - 4f_{i-1} + 3f_i\right)^2 \, ,
\end{equation}

\begin{equation}\label{beta1}
\beta_1 = \frac{13}{12}\left(f_{i-1} - 2f_i + f_{i+1}\right)^2 + \frac{1}{4}\left(f_{i-1} - f_{i+1}\right)^2 \, ,
\end{equation}
and
\begin{equation}\label{beta2}
\beta_2 = \frac{13}{12}\left(f_i - 2f_{i+1} + f_{i+2}\right)^2 + \frac{1}{4}\left(3f_i - 4f_{i+1} + f_{i+2}\right)^2 \, ,
\end{equation}
that measure the regularity of the k-th polynomial approximation $f^k_i$ at the stencil $S^k$.

Note that the choices of the coefficients in (\ref{f0}) - (\ref{f2}) and of the optimal weights $d_k$ follow the one in \cite{delzanna2007echo}, which differ from the one in the original paper, because it has been noted that these values suit better the high order scheme in combination with the derivation operation.

\subsection{Derivation operation}
The derivation operation is a high-order procedure which allows one to obtain a high order approximation from the point value quantities calculated at the intercell location.

This step has to be performed right after the computation of the fluxes via an approximate Riemann solver and it is necessary to preserve the accuracy in the calculation of spatial derivatives for schemes with order $n > 2$. As we did before, we will restrict the discussion to one dimension. The procedure described here follows the one outlined in the \texttt{ECHO} paper \cite{delzanna2007echo}.
Using this procedure, we will provide the numerical flux function $\hat{f}_{i+1/2}$, given a stencil of intercell fluxes $\{f_{i+1/2}\}$. 

The finite difference approximation of the first derivative in the point $x_i$ can be written as
\begin{eqnarray}
    h f'(x_i) &\approx \hat{f}_{i+1/2} - \hat{f}_{i-1/2} = \nonumber \\
    &= a( f_{i+1/2} - f_{i-1/2}) + b( f_{i+3/2} - f_{i-3/2}) + c(f_{i+5/2} - f_{i-5/2})
\end{eqnarray}
where the approximation has been truncated at sixth order and $h$ is the constant grid spacing.

If we now expand both sides of the equation in Taylor series around $x_i$ we find
\begin{equation}\label{der1}
	h f^{(1)}_i = \sum_{k=0}^{+\infty} f_i^{(k)} \frac{h^k}{k! 2^k} \left[ 1- (-1)^k\right]\left[a + 3^kb + 5^kc\right] \, ,
\end{equation}
where the exponents indicate the corresponding order of derivation, and the first derivative has been rewritten as $f_i^{(1)} \equiv f'(x_i)$. 
It is clear that all terms with even $k$ vanish. For $n=2$, where $b=c=0$, we find $a=1$. For $n=4$, where $c=0$, we have $a = 9/8$ and $b=-1/24$. Finally, for $n=6$, the solution is $a=75/64$, $b=-25/384$, $c=3/640$. The next step is to write
\begin{equation}\label{HO_approx}
	\hat{f}_{i+1/2} = d_0f_{i+1/2} + d_2 (f_{i-1/2} + f_{i+3/2}) + d_4 (f_{i-3/2} + f_{i+5/2}) \, ,
\end{equation}
and the comparison with (\ref{der1}) gives the relations $d_0 = a+b+c$, $d_2 = b+c$, $d_4 = c$.
The numerical values of $d_0$, $d_2$, and $d_4$ for the different order of approximation are provided in Table~\ref{coeffs_high_order}.
\begin{table}[t]
\caption{Coefficients of the approximation $\hat{f}_{j+1/2}$\, .}
\centering
    \begin{tabular}{|c|c|c|c|}
         \hline
         $n$ & $d_0$ & $d_2$ & $d_4$\\
         \hline
         2 & 1 & 0 & 0\\
         4 & 13/12 & -1/24 & 0\\
         6 & 1067/960 & -29/480 & 3/640\\
         \hline
    \end{tabular}
\label{coeffs_high_order}
\end{table}
Note that for $n=2$ one gets $\hat{f}_{j+1/2} = f_{j+1/2}$ as expected.

In order to highlight the nature of this procedure as a correction for higher than second order approximation, it is convenient to rewrite Equation (\ref{HO_approx}) as
\begin{equation}
    \hat{f}_{i+1/2} = f_{i+1/2} - \frac{1}{24} \Delta^{(2)} f_{i+1/2} + \frac{3}{640} \Delta^{(4)}f_{i+1/2}\, ,
\end{equation}
where only the first term is used in the case $n=2$, the second is added for $n=4$ and the complete expression is used for $n=6$. For a generic index $i$ the second and fourth order numerical derivative are given by
\begin{equation}
    \Delta^{(2)}f_i = f_{i-1} - 2f_i + f_{i+1}
\end{equation}
and
\begin{eqnarray}
\Delta^{(4)}f_i &= \Delta^{(2)}f_{i-1} - 2\Delta^{(2)}f_i + \Delta^{(2)}f_{i+1} = \nonumber \\
&= f_{i-2} - 4f_{i-1} + 6f_i - 4f_{i+1} + f_{i+2} \, ,
\end{eqnarray}
respectively.

\subsection{Simple Wave Test}
The first test performed to check the convergence of the total procedure is the evolution of a relativistic simple wave \cite{liang1977simplewave,anile1990relfluids}.
We have run this test using WENOZ as reconstruction method along with $n=2,4,6$ correction to the HLLE Riemann solver (in the following, they will be addressed as HLLE2, HLLE4, and HLLE6, respectively).

\begin{figure}[tp]
\centering
\begin{tabular}{cc}
\includegraphics[width=0.48\textwidth]{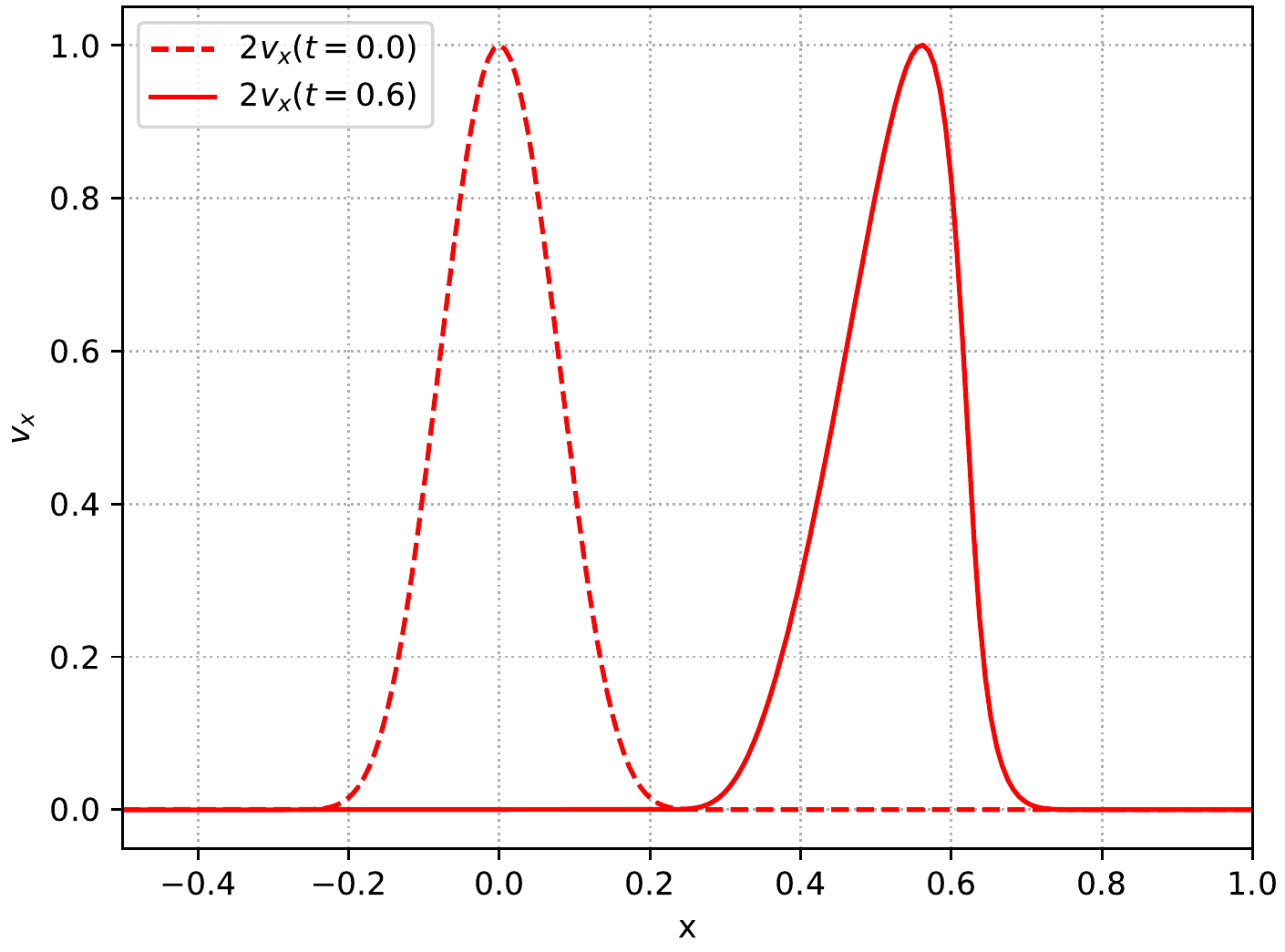} &
\includegraphics[width=0.48\textwidth]{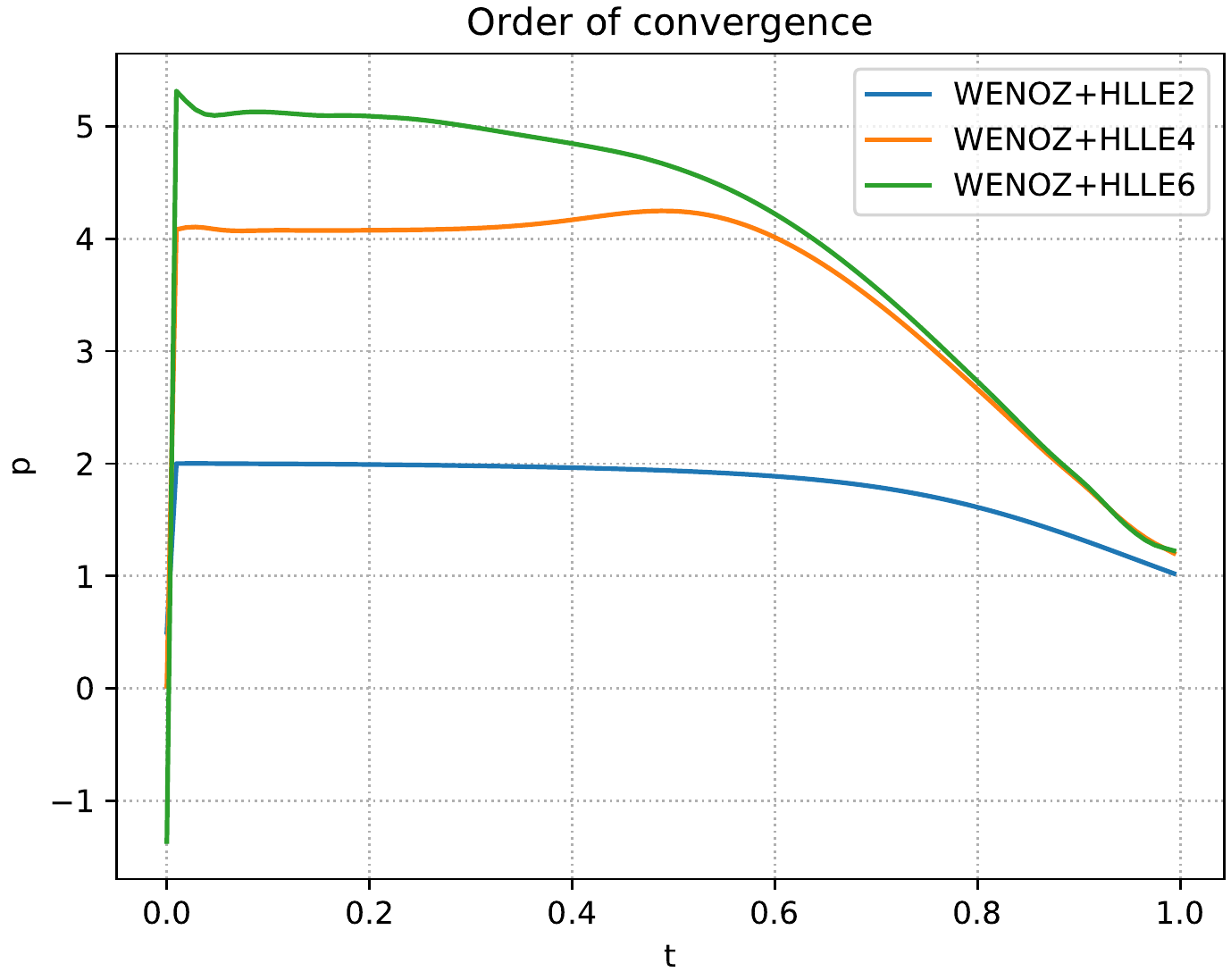} 
\end{tabular}

\caption{Left: Solution with 400 points. Right: Self-convergence factor (\ref{scf}), computed from three different resolutions: 400, 800, 1600 points.}
\label{SW_conv}
\end{figure}

The initial data are set up by choosing a reference state: following \cite{bernuzzi2016waveforms}, we chose a right-propagating simple wave with $\rho_0 = 1$ and $v_0 = 0$.
Assuming a polytropic EOS with $\Gamma = 5/3$ and $K = 100$, one can compute the sound speed in the reference frame via
\begin{equation}\label{c_0}
    c_0^2 = \frac{K \Gamma (\Gamma - 1) \rho_0^{\Gamma}}{(\Gamma - 1)\rho_0 + K \Gamma \rho_0^{\Gamma}}\,
\end{equation}
obtaining, in the specific case, $c_0 \approx 0.815$.
After the reference state has been defined, the velocity is perturbed with a sin-like function, so that its profile becomes (dashed line in the left panel of Figure \ref{SW_conv})
\begin{equation}\label{vel_pert}
    v = a \Theta(X - |x|) \sin^6 \left[\frac{\pi}{2} \left(\frac{x}{X} - 1\right)\right]\, ,
\end{equation}
where $\Theta(x)$ is the Heaviside function, $a = 0.5$, and $X = 0.3$.
Finally, the new sound speed is computed according to the Riemann invariant \cite{anile1990relfluids}
\begin{equation}
    c_s = \sqrt{\Gamma -1} \frac{\frac{\sqrt{\Gamma - 1} + c_0}{\sqrt{\Gamma - 1} - c_0} \left(\frac{1+v}{1-v}\right)^{\sqrt{\Gamma-1}/2} - 1}
    {\frac{\sqrt{\Gamma - 1} + c_0}{\sqrt{\Gamma - 1} - c_0} \left(\frac{1+v}{1-v}\right)^{\sqrt{\Gamma-1}/2} + 1}\, ,
\end{equation}
so that $c_s = c_0$ at $v=0$ and $c_s \rightarrow \sqrt{\Gamma - 1}$ as $v \rightarrow 1$.
The other quantities follow from the EOS: 
\begin{equation}
    \hat{\varepsilon} = \frac{c_s^2}{\Gamma (\Gamma - 1 -c_s^2)}\, ,
\end{equation}
\begin{equation}
    \hat{\rho} = \varepsilon^{1/(\Gamma - 1)}\, ,
\end{equation}
\begin{equation}
    \hat{p} = \varepsilon^{\Gamma / (\Gamma - 1)}\, ,
\end{equation}
where $\hat{\varepsilon}$, $\hat{\rho}$, and $\hat{p}$ are, respectively, the specific internal energy, the density, and the pressure normalized over the corresponding quantities in the reference state.
The solutions are computed on a 1-dimensional domain $[-1.5,1.5]$, employing  RK4 integrator for HLLE2 and HLLE4, and RK65 for HLLE6\footnote{\label{Runge-Kutta}This choice has been carried out in order to avoid a possible limitation on the order of convergence due to the Runge-Kutta integrator.}, with a CFL factor of $0.125$.

During the evolution, the profile of the wave begins to steepen until a shock is formed at $t \approx 0.63$ (see \cite{liang1977simplewave}).
In order to quantify the convergence properties of the various methods, we computed the self convergence factor defined as
\begin{equation}\label{scf}
    p \equiv \log_2 \left(\frac{||f(4\Delta x) - f(2\Delta x)||}{|| f(2\Delta x) - f(\Delta x)||}\right).
\end{equation}
The functions $f(\Delta x)$, $f(2\Delta x)$, and $f(4\Delta x)$ represent the numerical solutions calculated on uniform grids with corresponding grid spacing, and the norm employed is the L2-norm.
In this test, the three different resolutions are $\Delta x = 0.0075, 0.00375, 0.001875$, corresponding to 400, 800, 1600 points.

As it can be seen from the right panel of Figure (\ref{SW_conv}), the nominal convergence order is reached until the appearance of the shock. 
For both WENOZ+HLLE2 and WENOZ+HLLE4, the convergence is dominated by the order of the derivation operation; otherwise, for WENOZ+HLLE6, the convergence is dominated by the order of the reconstruction method and this is why we cannot get an order of convergence higher than fifth. As expected, the convergence order goes down for all methods when the shock is formed.

\subsection{Non-magnetized TOV}
A second test that has been performed is the evolution of a non-magnetized TOV star; the setup is the same used in the first paper of \texttt{Spritz} \cite{cipolletta2020spritz}. 
In particular, the initial configuration is generated using a polytropic EOS with $\Gamma = 2.0$ and $K = 100$, and initial rest-mass density $\rho = 1.28 \times 10^{-3}$.
The evolution of the system is then carried out adopting an ideal fluid EOS with the same value of $\Gamma$.
The physical domain is $[-20, 20]$ for $x$-, $y$-, and $z$-coordinates, with low, medium, and high resolution having $32^3$, $64^3$, and $128^3$ cells, respectively.
All the tests lasted for 5 ms using the WENOZ reconstruction method and the three approximation for the Riemann solver (HLLE2, HLLE4, and HLLE6). In the cases of HLLE2 and HLLE4, RK4 method is employed for time stepping, while RK65 is used in HLLE6 case, with a CFL factor of $0.25$.

\begin{figure}[tp]
\centering
\begin{tabular}{cc}
\includegraphics[width=0.49\textwidth]{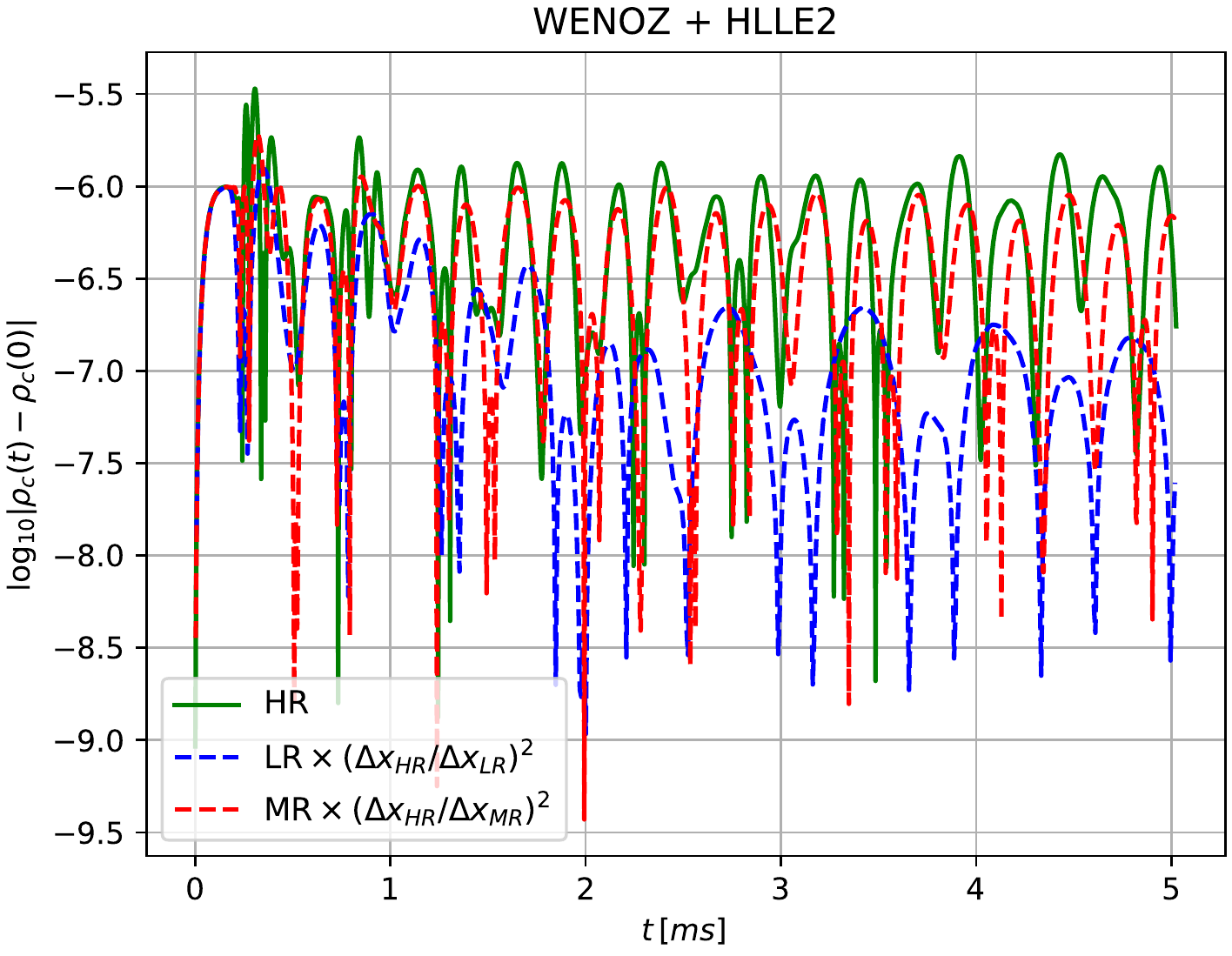} &
\includegraphics[width=0.49\textwidth]{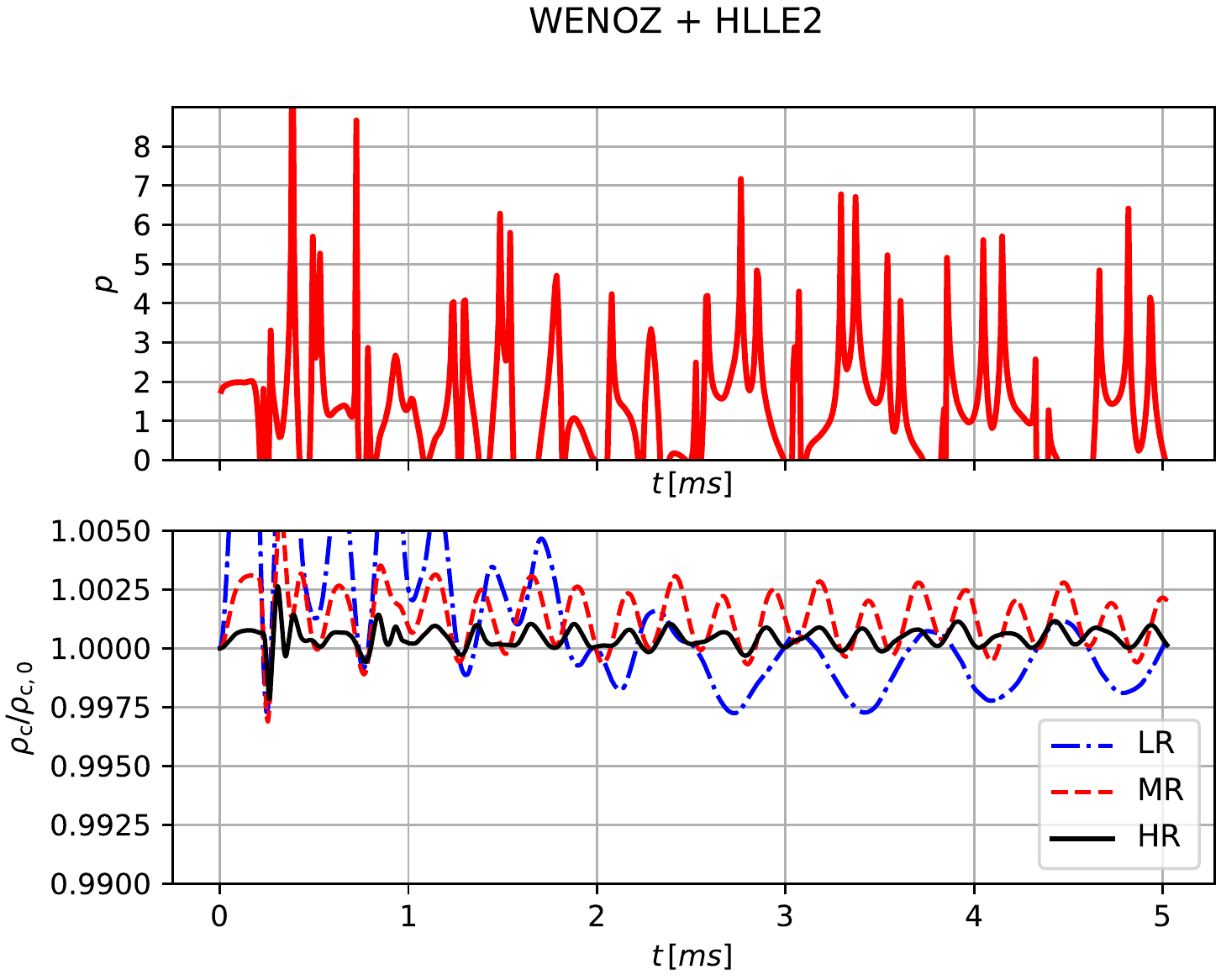} \\
\includegraphics[width=0.49\textwidth]{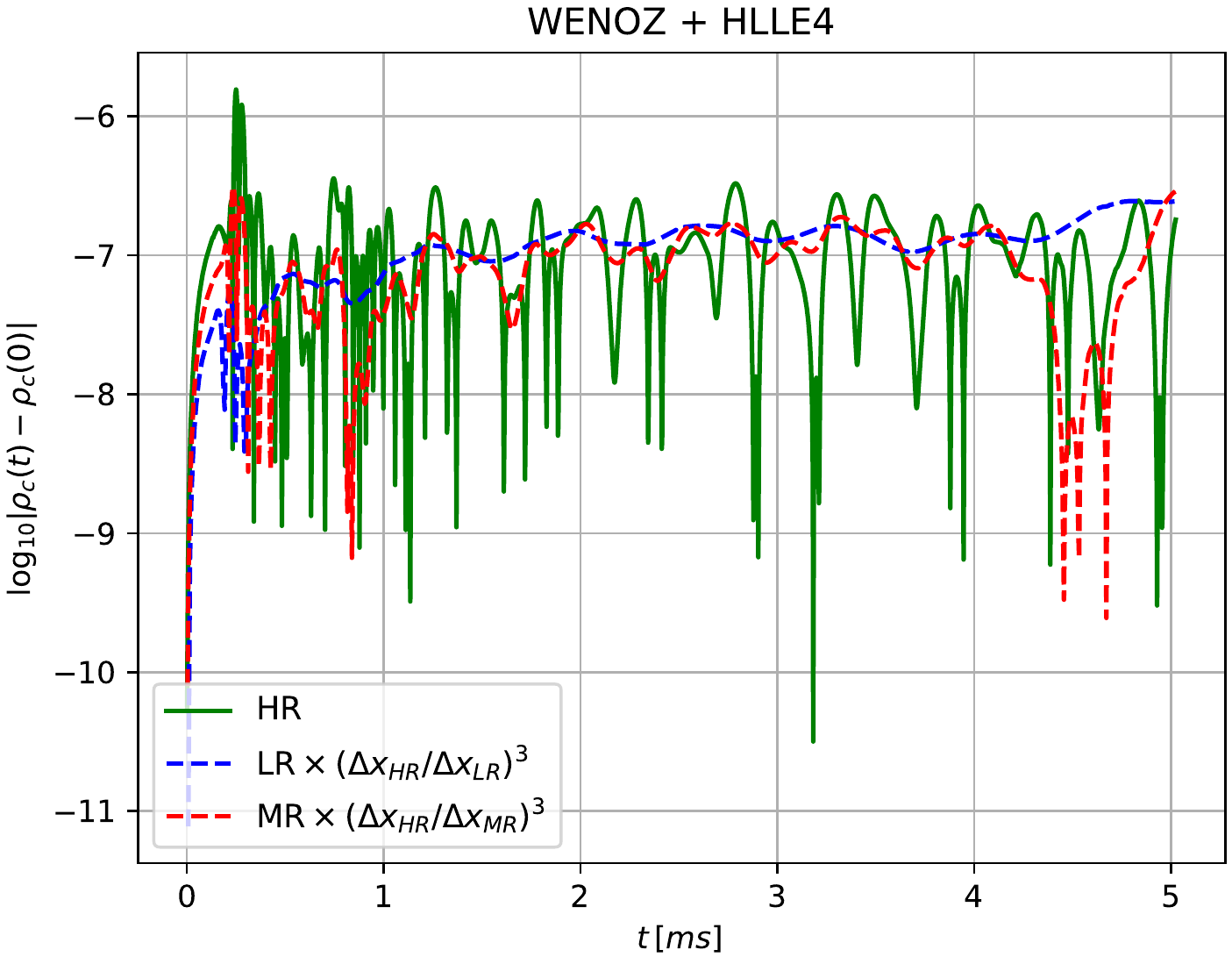} &
\includegraphics[width=0.49\textwidth]{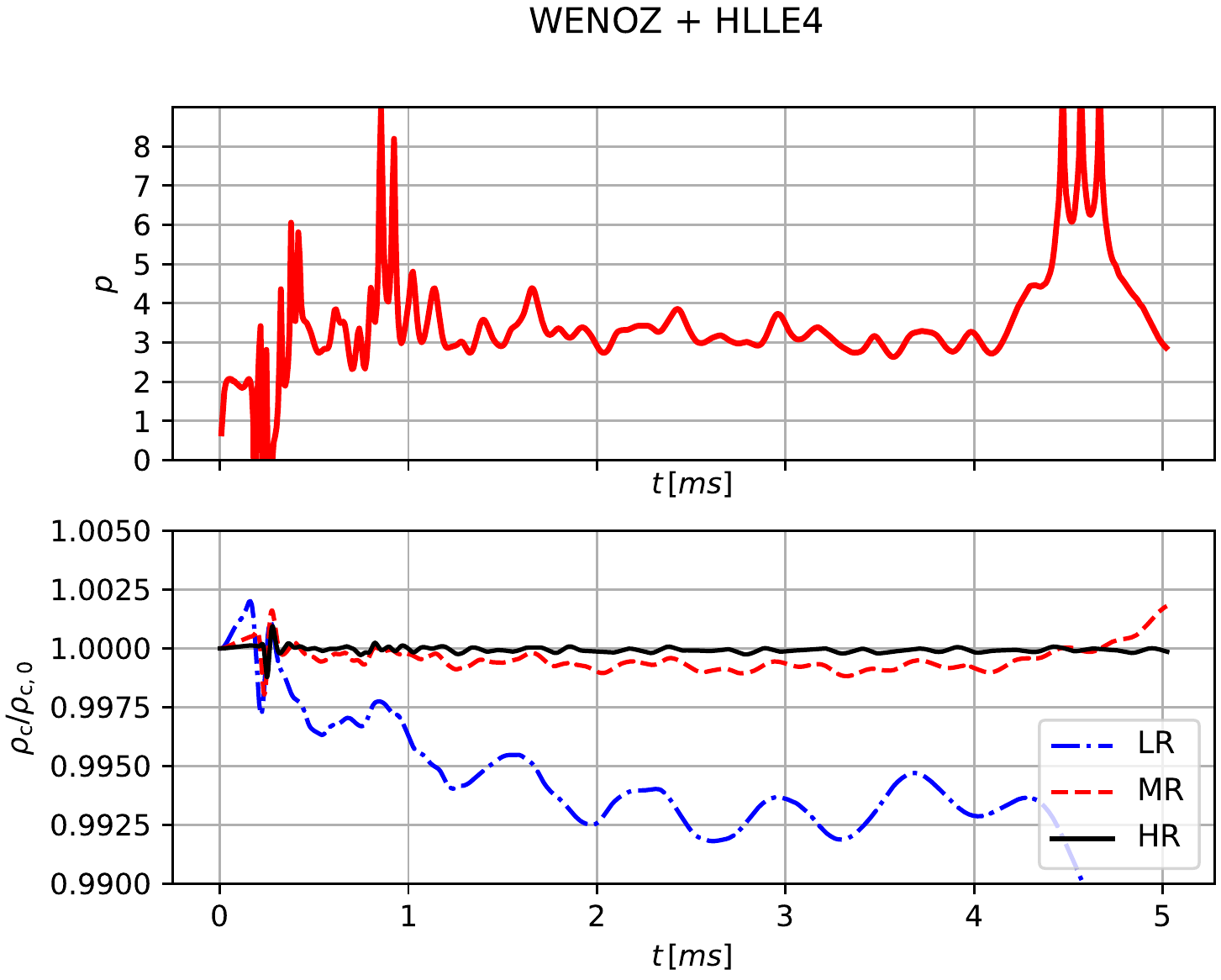} \\
\includegraphics[width=0.49\textwidth]{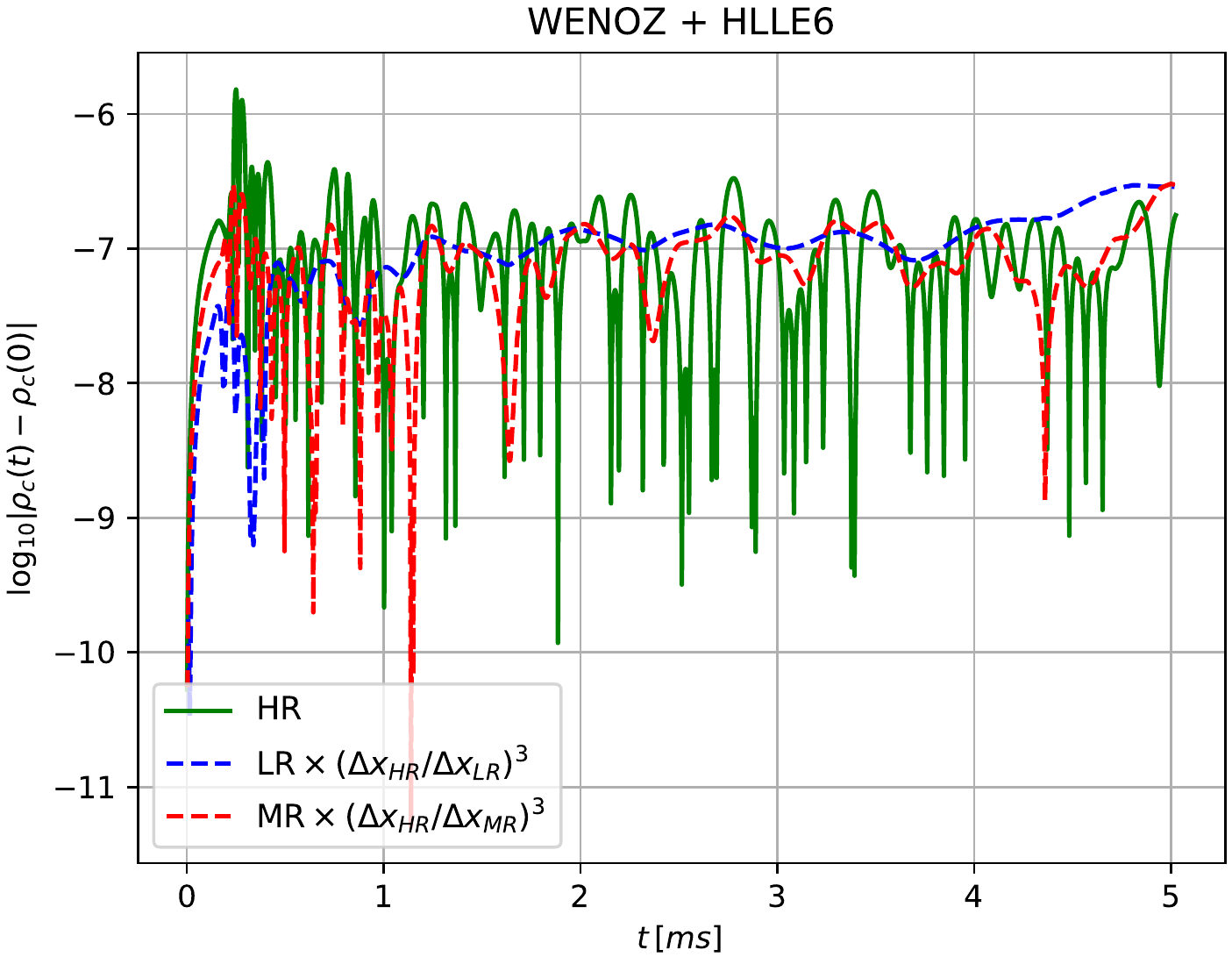} &
\includegraphics[width=0.49\textwidth]{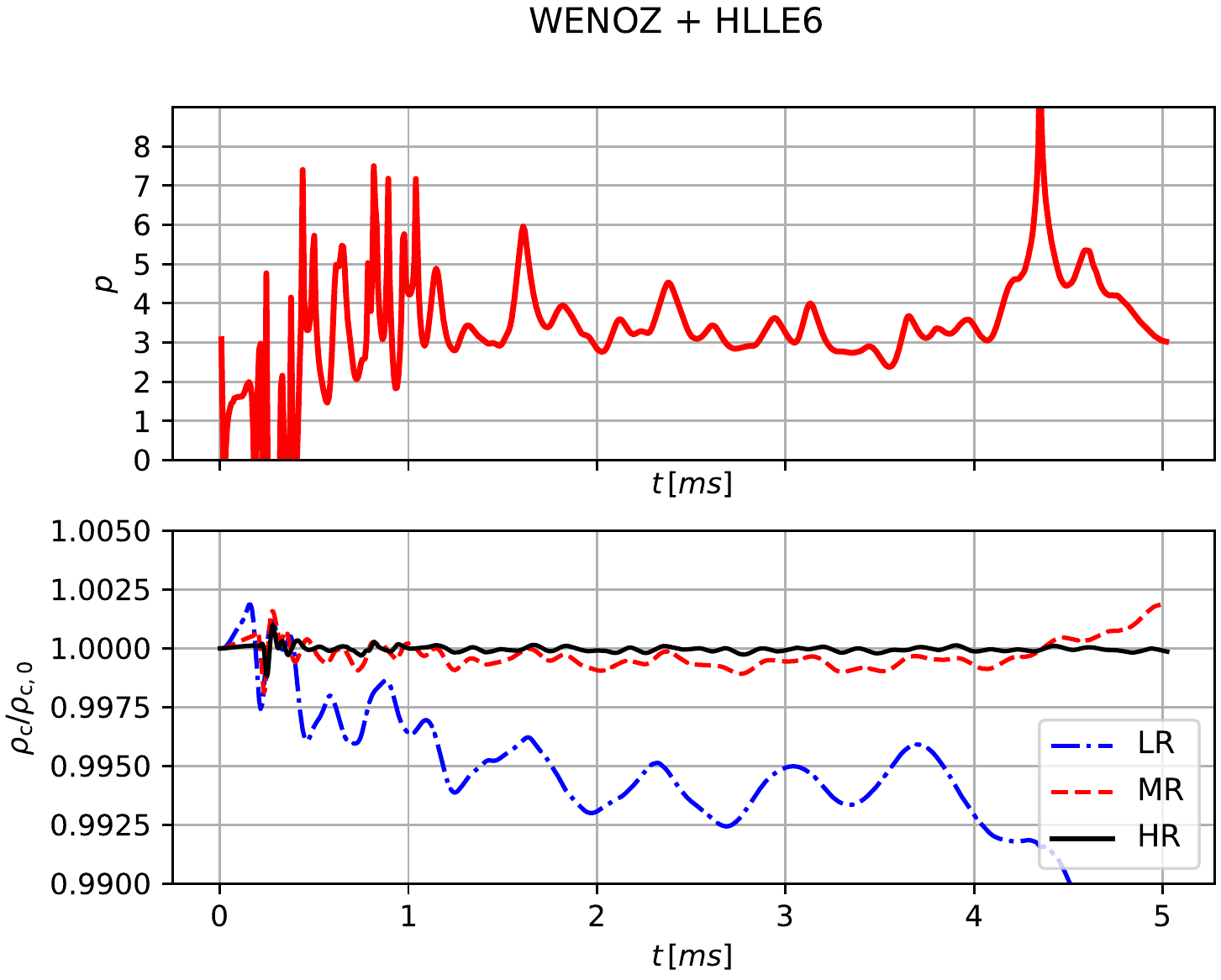}
\end{tabular}
\caption{Left: Evolution of $|\rho_c(t) - \rho_c(0)|$. Right: Self-convergence factor $p$ (top panels), computed from the three different resolutions ($32^3$, $64^3$, $128^3$ points), and $\rho_{\rm max}/\rho_{{\rm max}, 0}$ (bottom panels).}
\label{TOV}
\end{figure}

In the continuum limit, the evolution of this kind of system is trivial; however, the discretization of the problem brings errors (due to the discretization itself) that cause radial oscillations, which are observable, for example, in the central rest-mass density (see Fig.~\ref{TOV}). 
The amplitude of these oscillations becomes smaller as the number of points increases.
In the right panels of Figure (\ref{TOV}) it is possible to note that the density has a peculiar behavior for low and medium resolution at late times. 
This fact can be traced back to the choice of the ideal fluids EOS in the evolution of the system: it is known that truncation errors with this EOS are very large, because significant unphysical shock-heating is observed at low densities \cite{bernuzzi2016waveforms}.

In order to verify the convergence of the high-order methods, we compute the self-convergence factor (based on deviations of central rest-mass density with respect to the initial value), which oscillates around the value $p = 3$ for both HLLE4 and HLLE6.
Such order of convergence is maintained until the aforementioned truncation errors become significant, i.e., until $\sim 4$ ms.

\begin{figure}[tp]
    \centering
    \includegraphics[scale=0.6]{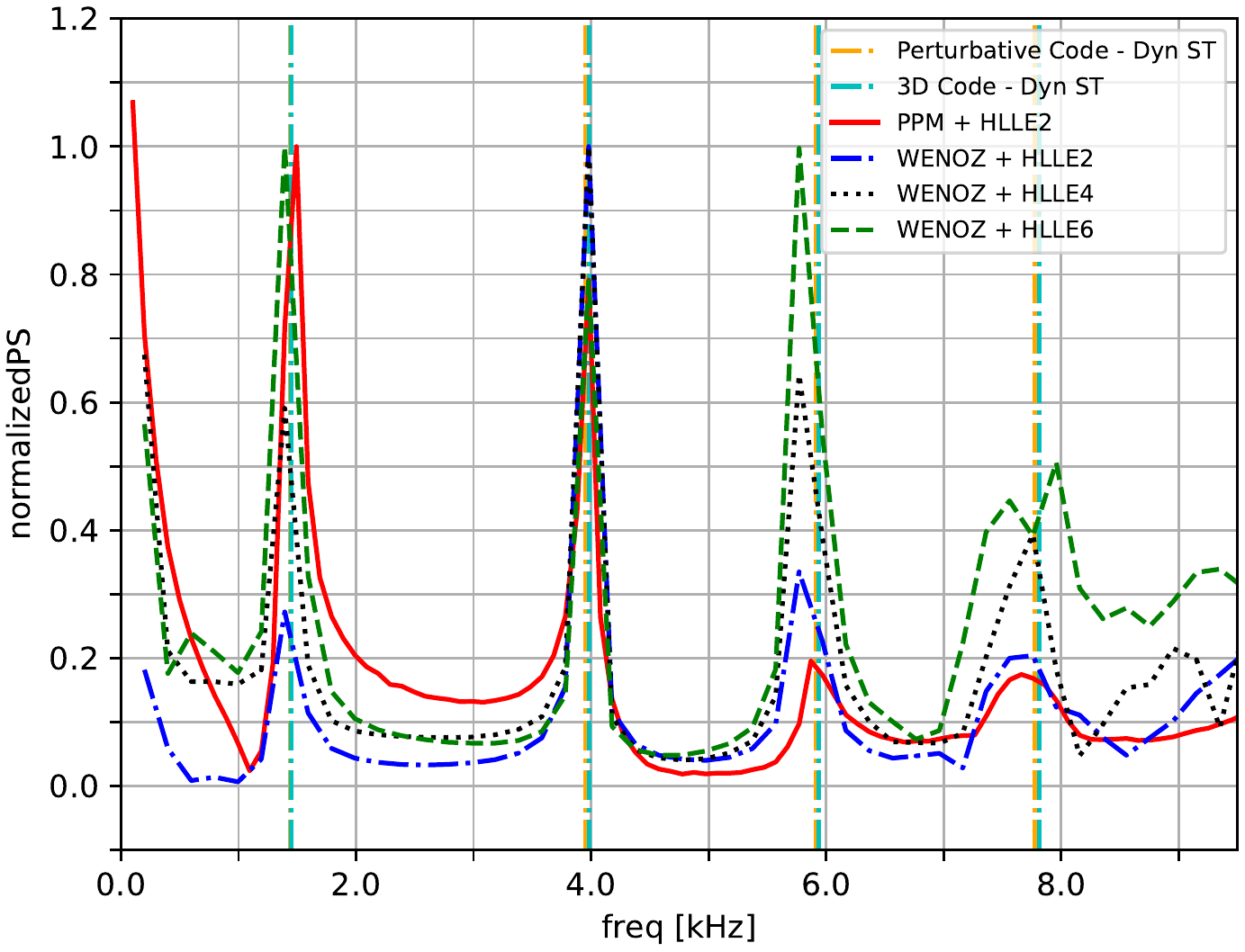}
    \caption{Power spectrum of the central rest-mass density evolution, normalized to the maximum amplitude of the oscillation frequency peaks.}
    \label{TOV_ps}
\end{figure}

In the end, Figure (\ref{TOV_ps}) reports the power spectrum of the evolution of the rest-mass densities of the different runs.
The power spectrum is computed via a fast Fourier transform (FFT) in order to extract the amplitudes and the frequencies of the oscillations, and then the amplitudes are normalized to the maximum for each simulation.
Figure (\ref{TOV_ps}) also shows the peaks' frequencies of the oscillations taken from the literature \cite{Font2001}, that were obtained with independent codes.
All the simulations show a good agreement with each other and the independent results. In particular, it is worth noting that the high-order reconstruction coupled to high-order Riemann solvers (black-dotted and green-dashed curves in the figure) is evidently capable of better resolving the overtones (i.e.~the higher frequency peaks in the spectrum) with respect to the lower-order methods (red-solid and blue-dash-dotted curves in the figure).

\subsection{Magnetized TOV with Tabulated EOS}

Finally, we performed a test evolving a magnetized TOV star using a tabulated EOS (LS220), with a $S$-slicing initial condition, and employing WENOZ as reconstrunction method and the 4-th order approximation for the HLLE Riemann solver (this case has been called WENOZ+HLLE4).
This case is then compared with the same case evolved using PPM reconstruction method and the 2-nd order approximation to HLLE, denoted with PPM+HLLE2.

\begin{figure}[tbh!]
    \centering
    \begin{tabular}{ccc}
        \includegraphics[width=0.45\textwidth]{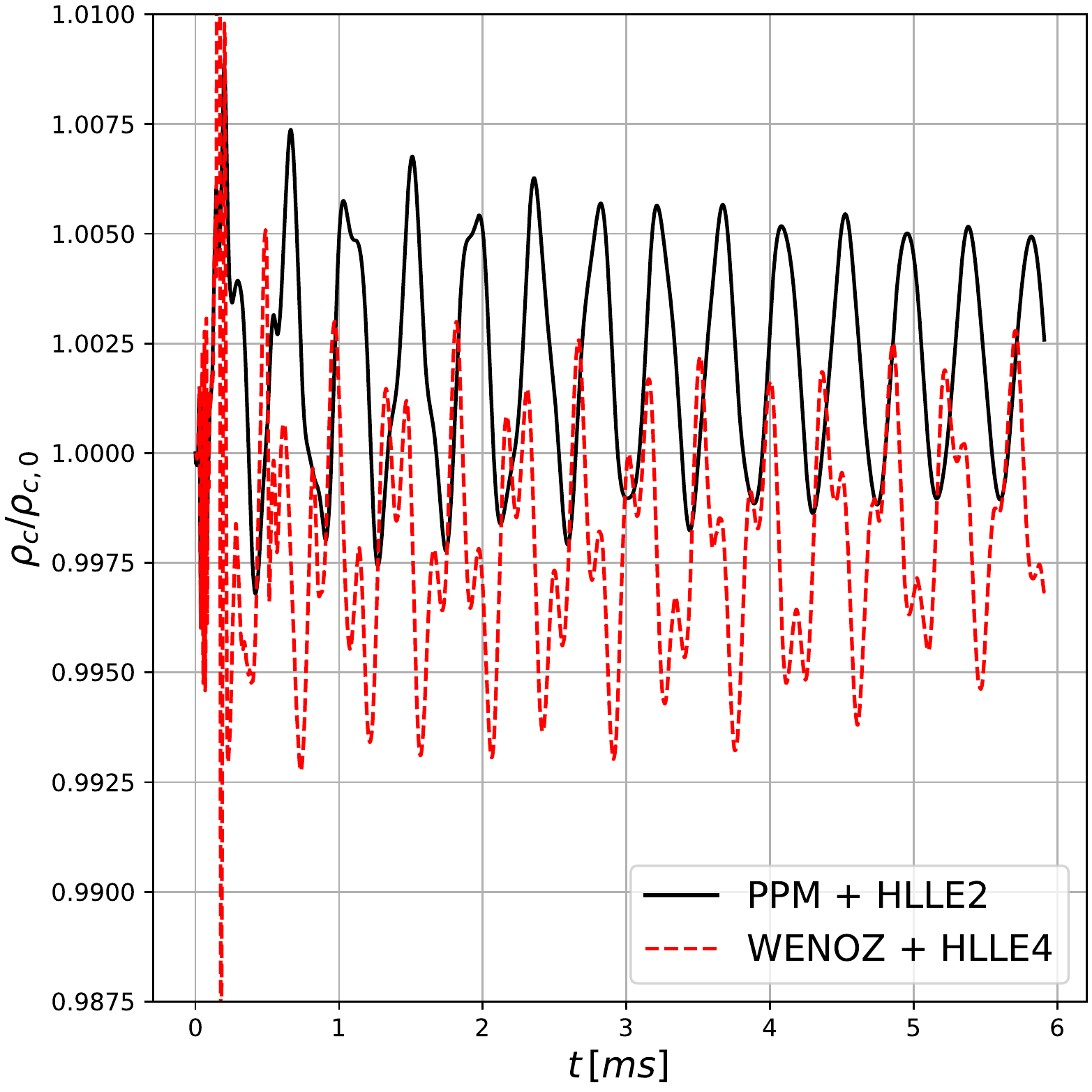}  &
        \includegraphics[width=0.45\textwidth]{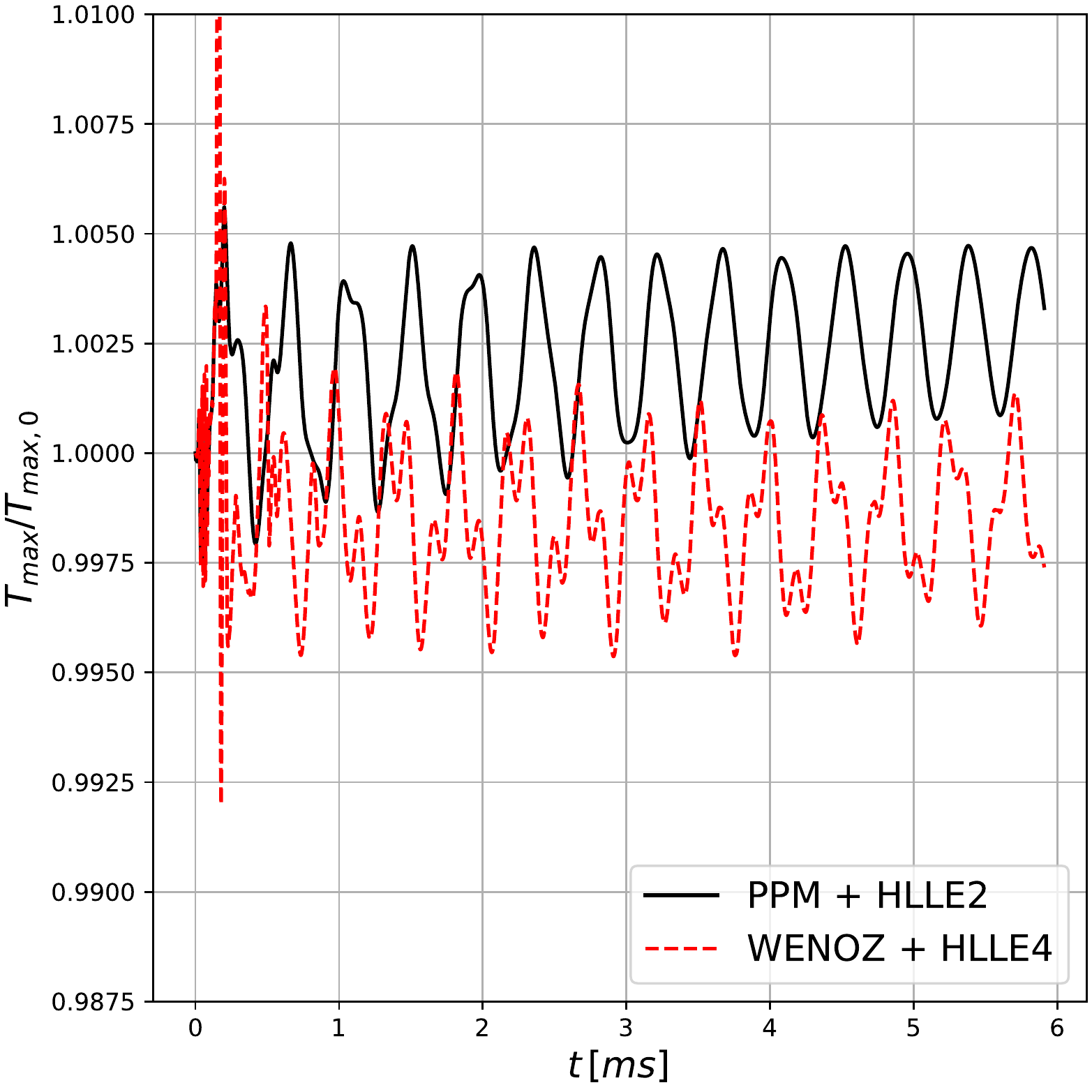} &
    \end{tabular}
    \includegraphics[width=0.45\textwidth]{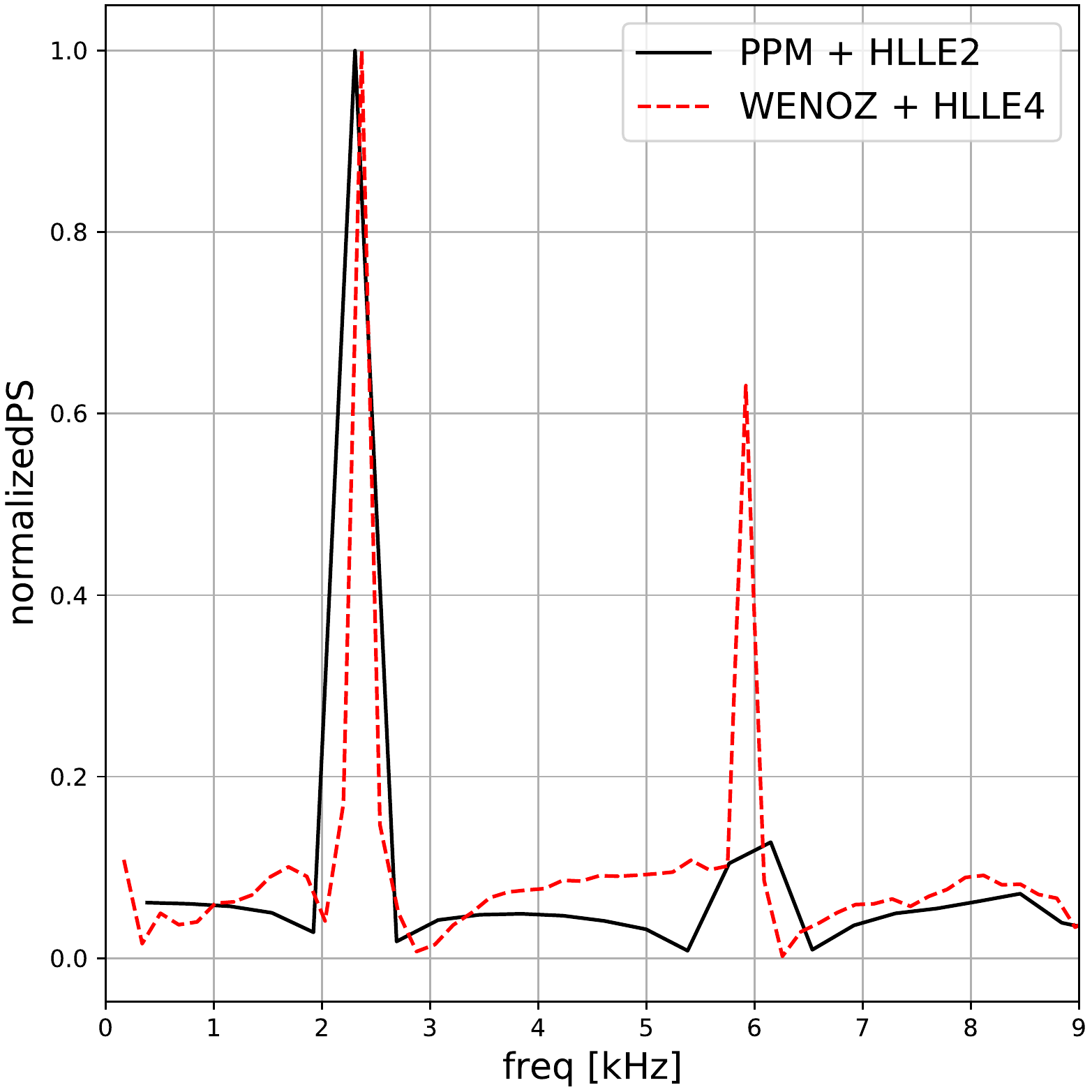}
    \caption{Comparison between the results obtained with and without high-order methods for a TOV evolved with the LS220 EOS. 
    Upper panels: Evolution of the central rest-mass density (left) and the maximum of the temperature (right), both normalized to their initial values. 
    Bottom panel: Normalized power spectrum of the central rest-mass density.}
    \label{sim14_HO}
\end{figure}

The upper panels of Figure (\ref{sim14_HO}) show the evolution of the central rest-mass density $\rho_c$ and of the maximum of the temperature $T_{max}$, both normalized over their initial values, respectively $\rho_{c,0}$ and $T_{max,0}$.
The results obtained with the use of the high-order scheme, in particular the setup WENOZ+HLLE4, are more precise than the ones obtained with the older version of the \texttt{Spritz} code; using high-order methods helps reducing the oscillations around the real value.
Moreover, enabling WENOZ and the fourth-order correction to HLLE softens the slight increasing behaviour of $T_{max}$, as shown in the upper right panel of Figure (\ref{sim14_HO}).

The gain in accuracy is particularly evident in the plot for the power spectrum of the evolution of the central rest-mass density, shown in the lower panel of Figure (\ref{sim14_HO}).
Each power spectrum is computed, as before, via the FFT and the amplitudes are normalized over their maximum for each simulation.
It can be easily seen that, while the lower-order version of \texttt{Spritz} shows a noticeable peak only for the fundamental frequency, the high-order upgrade can resolve very well also the first overtone, which results to be more prominent than the one of the PPM+HLLE2 case.

\newpage

\section*{References}
\bibliographystyle{unsrt}
\bibliography{references}

\end{document}